\documentclass{iopjournal}
\usepackage[english]{babel}
\usepackage[utf8]{inputenc}
\usepackage{dcolumn}
\usepackage{bm}
\usepackage{graphicx}
\usepackage{amsmath,amsfonts,amssymb}
\usepackage{enumitem}
\usepackage{latexsym}
\usepackage{array}
\usepackage{epsfig}
\usepackage{color}
\usepackage[T1]{fontenc}
\usepackage[sc]{mathpazo}
\usepackage{siunitx}
 \usepackage{etoolbox}
\usepackage[
    backend=biber,
    style=phys,           % Uses the specialized physics style
    articletitle=false,   % Strips out the article titles
    biblabel=brackets,    % Formats labels as [1] instead of 1.
    giveninits=true       % Forces first names to be initials
]{biblatex}
\usepackage{varioref}
\labelformat{figure}{Fig.\;#1}

\newcommand{\nocontentsline}[3]{}
\newcommand{\tocless}[2]{\bgroup\let\addcontentsline=\nocontentsline#1{#2}\egroup}

\makeatletter
\def\maketitle{
\@author@finish
\title@column\titleblock@produce
\suppressfloats[t]}
\makeatother

\newcommand{\aref}[1]{\hyperref[#1]{Appendix~\ref*{#1}}}

\newcommand{\eq}{Eq.\,}

\newcommand{\cf} {cf.~}

\newcommand{\eg} {e.g.~}
\newcommand{\rref} {Ref.\,}

\DeclareMathOperator{\Hc}{H.c.}

\newcommand{\Jt} {{\cal J}_{k}}
\newcommand{\R} {{\bf R}}
\DeclareMathOperator{\hc}{H.c.}

\begin{document}

\articletype{PAPER}

\title{Emergent cavity-QED dynamics along the edge of a photonic lattice}
	
\author{Enrico Di Benedetto$^{1,*}$, Xuejian Sun$^{1,2}$, Marcel A. Pinto$^1$, Luca Leonforte$^1$, Chih-Ying Chang$^{3,4}$, Vincent Jouanny$^{3,4}$, Léo Peyruchat$^{3,4}$, Pasquale Scarlino$^{3,4}$ and Francesco Ciccarello$^{1,5}$}

\affil{$^1$Universit{\`a} degli Studi di Palermo, Dipartimento di Fisica e Chimica -- Emilio Segr{\`e}, via Archirafi 36, I-90123 Palermo, Italy}

\affil{$^2$School of Physics and Telecommunication Engineering, Zhoukou Normal University, 466001 Zhoukou, China}

\affil{$^3$Hybrid Quantum Circuits Laboratory (HQC), Institute of Physics, \'{E}cole Polytechnique F\'{e}d\'{e}rale de Lausanne (EPFL), 1015, Lausanne, Switzerland}

\affil{$^4$Center for Quantum Science and Engineering, Institute of Physics, \'{E}cole Polytechnique F\'{e}d\'{e}rale de Lausanne (EPFL), 1015, Lausanne, Switzerland}

\affil{$^5$NEST, Istituto Nanoscienze-CNR, Piazza S. Silvestro 12, 56127 Pisa (Italy)}

\affil{$^*$Author to whom any correspondence should be addressed.}

\email{enrico.dibenedetto@unipa.it (corresponding author)}

\keywords{circuit-QED, cavity-QED, flat band physics, Jaynes-Cummings model, power-law localization, photonic lattices, honeycomb lattice}
 
\begin{abstract}
We investigate qubits coupled to the boundary of a two-dimensional photonic lattice that supports {\it dispersionless} edge modes -- unlike conventional edge modes that sustain propagating photons.
As a case study, we consider a honeycomb lattice (\textit{photonic graphene}) of coupled resonators with a zigzag edge, where the edge modes form a flat band defined only over a restricted region of momentum space.
We show that light–matter interactions are effectively captured by a dissipative cavity-QED model, wherein the emitter coherently couples to a localized mode emerging as a superposition of edge modes.
This mode has support on only one sublattice and displays an unconventional power-law localization around the qubit -- yet remaining normalizable in the thermodynamic limit -- with a spatial range that can be tuned by introducing lattice anisotropy.
We predict the occurrence of vacuum Rabi oscillations and state transfer between distant emitters.
An experimental demonstration using superconducting circuits is proposed.
\end{abstract}
	
\section{Introduction}

\begin{figure}[t]
    \centering 
    \includegraphics[width=0.8\textwidth]{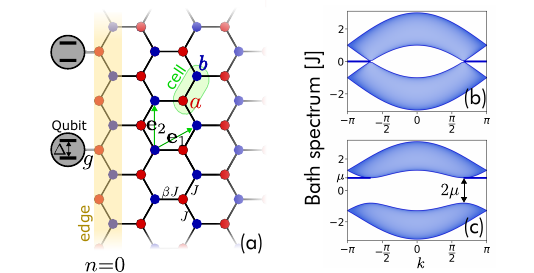}
    \caption{(a) Qubits (two-level emitters) coupled to a zigzag edge of a 2D honeycomb lattice of coupled resonators (\textit{photonic graphene}).
The bare frequency of $a/b$ resonators is $\pm \mu$ (not shown).
(b)-(c) Spectrum of bath normal modes in the gapless case $\mu{=}0$ (b) and gapped case $\mu{>}0$ (c).
In (b) and (c), a set of dispersionless edge modes (partial flat band) arises at frequency zero and $\mu$, respectively.}
    \label{fig:fig1}
\end{figure}

Two-dimensional (2D) photonic lattices with non-trivial topology 
support chiral edge modes that spectrally arise within bandgaps of bulk modes \cite{kim2020recent,ozawa_topological_2019,weber20242024}.
Coupling quantum emitters to an edge of such a photonic bath and adjusting their frequency within a bandgap thus enables tasks such as directional emission \cite{barik2018topological,jalali2020chiral,owens2022chiral,vega_topological_2023,suarez2025chiral,jouanny2025superstrong,pakkiam2024experimental} and state transfer \cite{yao2013topologically,lemonde2019quantum,dlaska2017robust}.
In such dynamics, an excited emitter typically undergoes an {\it irreversible} decay into the photonic edge modes, with these modes embodying a 1D continuum characterized by an associated dispersion relation.
Here, we predict that light-matter interactions on the edge of a 2D photonic lattice can exhibit a different character, resembling instead {\it reversible} cavity-QED dynamics.
As a case study, we consider a set of qubits (two-level quantum emitters)  coupled to the edge of photonic \textit{graphene} \cite{Castro_Graphene_2009}, i.e., a semi-infinite 2D honeycomb lattice of coupled resonators.
For suitable edge geometry, e.g., the \textit{zigzag} boundary in \ref{fig:fig1}(a), graphene can host topological edge modes, albeit of an unusual type \cite{fujita1996peculiar,nakada1996edge, Castro_Graphene_2009, tan_edge_2021, kohmoto_zero_2007} in that [\cf \ref{fig:fig1}(b)]: {\it (i)} they are {\it dispersionless}, thus forming a 1D {\it flat band} \cite{leykam_artificial_2018} sustaining non-propagating photons;
{\it (ii)} they exist only within a {\it limited} region of the first Brillouin zone (BZ), a major consequence being that they cannot be expanded in a basis of spatially compact modes \cite{park2024quasi} unlike usual flat bands \cite{rhim_classification_2019};
{\it (iii)} spectrally they do not lie within a bandgap but rather touch both the bands of bulk modes at the Dirac points [see \ref{fig:fig1}(b)].
Our goal here is investigating light-matter interactions when qubits are tuned near the frequency of edge modes of this kind.
Spectrally-isolated flat bands coupled to an emitter reduce to an effective Jaynes-Cummings (JC) model with  the cavity mode embodied by an emergent superposition of flat-band eigenmodes, exponentially localized around the emitter \cite{de_bernardis_light-matter_2021,di2025dipole}.
Whether this picture applies here is non-trivial: despite the lack of spectral isolation [\cf\ref{fig:fig1}(b)], the dispersive bands exhibit a vanishing density of states (DOS) at the flat-band frequency, coinciding with the Dirac points.
\cite{Castro_Graphene_2009}. However, this vanishing DOS is accompanied by a singularity that, for emitters coupled to the lattice \textit{bulk}, was shown to trigger non-Markovian effects - notably, the formation of atom-photon \textit{quasi}-bound states that prevent full emitter decay \cite{gonzalez2018exotic,redondo2021quantum} if the lattice has a finite size.
Despite these complexities, we find a wide regime where the dynamics is effectively captured by a (generally dissipative) JC model based on a localized mode at the lattice edge emerging from a superposition of flat-band eigenmodes. This allows us to build an effective model in which the cavity mode inherits the scaling properties of the partial FB, in particular its power-law scaling with distance \cite{park2024quasi}.
Unlike standard flat bands, however, this mode exhibits {power-law} localization around the qubit.
In addition to the coherent coupling to this mode, the qubit undergoes dissipation into bulk modes (the dispersive bands).
The associated loss rate is however (linearly) dependent on the qubit’s detuning from the localized mode;
in particular, it vanishes at zero detuning, reflecting the zero DOS at the Dirac points.
Notably, unlike the bulk case \cite{gonzalez2018exotic,redondo2021quantum}, we show that atom-photon bound states due to the Dirac cone's singularity do not contribute to the dynamics here [see \aref{subsec:non-Markov}].
Using this model and its multi-qubit generalization, we predict the occurrence of vacuum Rabi oscillations and quantum state transfer \cite{diekmann_2026_enabling} along the lattice edge.

\section{Model}
Our photonic bath $B$ is a 2D honeycomb lattice of coupled single-mode resonators [see \ref{fig:fig1}(a)] whose Hamiltonian reads (throughout we set $\hbar=1$)
\begin{equation}
    \label{eq:HB}
    H_B = \mu\sum_{\R}(a_{\R}^\dagger a_{\R}- b_{\R}^{\dagger}b_{\R})+
\sum_{\langle\R,\R'\rangle} J_{\langle\R,\R'\rangle}(a_{\R}^\dagger b_{\R'} + \Hc),
\end{equation}
where $\R= n{\bf e}_1 + m{\bf e}_2$ specifies the unit cell position (with ${\bf e}_{1(2)}$ primitive vectors and $n,m$ integers), while $a_{\R}\equiv a_{n m}$ and $b_{\R}\equiv b_{n m}$ are standard bosonic ladder operators (here $\langle\R,\R'\rangle$ denotes nearest-neighbor resonators).
In \eq\eqref{eq:HB}, $\pm \mu$ is the bare frequency of $a$- and $b$-resonators, while $J_{\langle\R,\R'\rangle}$ is the photon hopping rate [see \ref{fig:fig1}(a)].
Unless otherwise stated, we will set $J_{\langle\R,\R'\rangle}=J$ [corresponding to $\beta{=}1$ in \ref{fig:fig1}(a)].
For $\mu=0$, \eq\eqref{eq:HB} is the bosonic analogue of the celebrated tight-binding Hamiltonian of graphene \cite{Castro_Graphene_2009}.
Importantly, as shown in \ref{fig:fig1}(a), here we consider a {\it semi-infinite} lattice with a zigzag {\it edge} (resonators $a_{n=0\,m}$ and $b_{n=0\,m}$).
Accordingly, $n= 0,1,2$,..., while $m\in \mathbb{Z}$.
 
 A set of qubits (two-level emitters), each with ground (excited) state $|g_j\rangle$ ($|e_j\rangle$) whose energy separation is $\Delta$, are locally weakly--coupled under the rotating-wave approximation to $a$-resonators [see \ref{fig:fig1}(a)].
The total Hamiltonian thus reads
 \begin{equation}
    H=\Delta \sum_{j} \sigma_j^\dagger \sigma_j+H_B+g\sum_{j}\left(\, a_{\R_j}^{\dagger}\sigma_{j}+\Hc\right)\,, 	\label{Htot}
 \end{equation}
with $\sigma_j = |g_j\rangle\langle e_j|$ the pseudo-spin ladder operator of the $j$th qubit, $g$ the atom-photon coupling strength and where ${\R}_j\equiv (0,m_j)$ labels the edge resonator which qubit $j$ is coupled to.
\section{Bulk and edge modes of $H_B$}
By enforcing periodic boundary conditions {\it along} the edge direction ${\bf e}_2$ [\cf\ref{fig:fig1}(a)], whose corresponding momentum component is called $k$, the bath Hamiltonian \eqref{eq:HB} can be exactly mapped to a set of uncoupled 1D Rice-Mele models \cite{asboth2016short}, each labeled by $k$ and having an edge at $n=0$ (see \eg \rref\cite{tan_edge_2021}).
This allows to express $H_B$ in the diagonal form [see \aref{app:geometry}]
\begin{equation}
    H_B= \sum_{\nu=\pm}\int_{0}^\pi \mathrm{d} q\int_{-\pi}^\pi\!\mathrm{d} k \,\omega_{\nu}(k,q) {\cal B}^\dagger_{\nu}(k,q){\cal B}_{\nu}(k,q)+\mu\int\limits_{2\pi/3 <|k|\leq \pi}\!\mathrm{d} k  \, {\cal E}^\dagger_{k} \mathcal{E}_k
\end{equation}
with $q$ the momentum component corresponding to direction ${\bf e}_1$ (from edge to bulk).
Two symmetric bands of bulk modes ${\cal B}_{\pm}(k,q)$ arise. For $\mu=0$, their spectra $\omega_{\pm}(k,q)$ [see \ref{fig:fig1}(b)-(c)] are the same as those for bulk graphene \cite{Castro_Graphene_2009}.
These bands touch at the Dirac points ($q=\pi$, $k=\pm 2\pi/3$), while for $\mu>0$ a gap of width $2\mu$ opens up.
Besides the bands of bulk modes, the lattice additionally hosts a set of {\it  dispersionless edge modes} ${\cal E}_k$ with common frequency $\mu$, where, importantly, $k$ takes only values external to the two Dirac points.
Modes ${\cal E}_k$ thus form a {\it partial} flat band [\cf \ref{fig:fig1}(b)-(c)].
Their real-space representation is $\mathcal{E}_k=\sum_{n,m} \varepsilon_k (n,m) a_{n m}$, where
\begin{equation}
    \label{eq:edge-states}
    \varepsilon_k (n,m)= \frac{\mathcal{N}_{k}}{\sqrt{2\pi}} (-1)^n e^{-ik(m+n/2)}\, e^{-n/\lambda_k}\,,
\end{equation}
with $\mathcal{N}_k = \sqrt{-(1+2\cos{k})}$ and $\lambda_k^{-1}= \ln\left|{J/{\cal J}_{k}}\right|$ for $\Jt=2J\cos{(k/2)}$.
Edge modes ${\cal E}_k$ thus have support only on the $a$-sublattice, being localized on the edge $n=0$ with a penetration length $\lambda_k$ ranging between 0 (for $k=\pm\pi$) and infinity (for $k=\pm2\pi/3$).
Their emergence for $\mu=0$ can indeed be linked \cite{ryu2002topological} to well-known zero-energy edge states of the standard SSH model in the non-trivial phase protected by sublattice, i.e., chiral, symmetry \cite{asboth2016short}.
For $\mu>0$ (open gap), chiral symmetry is lost, but edge modes are unaffected with their frequency now shifted by $\mu$ \cite{ryu2002topological}.
\begin{figure}
    \centering   
    \includegraphics[width=0.8\textwidth]{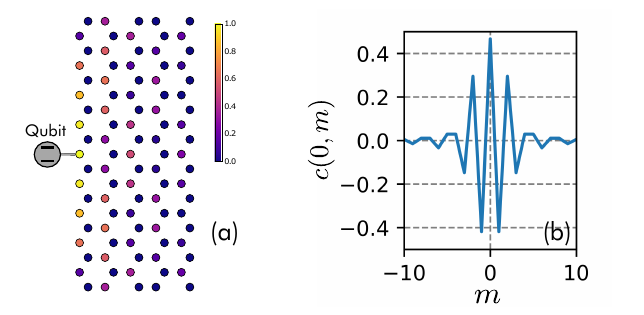}
    \caption{ (a) Real-space distribution (modulus) of mode ${\cal C}$ for one qubit coupled to $a_{00}$ (we used a logarithmic color scale and rescaled to the maximum value).
(b) Mode amplitude $c(0,m)$ along the edge [\cf\eq\eqref{eq:c0m}].}
    \label{fig:fig2a}
\end{figure}

\section{ Effective cavity-QED model for $\mu=0$: one qubit}
We first set $\mu=0$ [\cf \ref{fig:fig1}(b)] and consider only one qubit [thus we drop subscript $j$ in \eqref{Htot}] coupled to the resonator $a_{00}$.
We expand $a_{00}$ in terms of modes ${\cal E}_k$ (having zero frequency) and ${\cal B}_\pm (k,q)$ so as to split the coupling Hamiltonian [last term in \eq\eqref{Htot}] into one contribution from edge modes and one from bulk modes.
The former can be recast as $V_{\rm FB} = g ({\cal C}^\dag \sigma+\Hc)$ with ${\cal C}=\sum_k \varepsilon^*_k (0,0) {\cal E}_k$ a superposition mode which is itself a zero-frequency normal mode of $H_B$ whose detuning  from the qubit is measured by $\Delta$.
For $g$ and $|\Delta|$ small enough compared to $J$, this contribution will dominate over the one from bulk modes, whose DOS vanishes on approaching the Dirac points \cite{Girvin_Yang_2019}.
By treating these as a Markovian bath \cite{breuer2002theory}, we thus end up with a master equation for the joint density matrix $\rho$ of the qubit and mode ${\cal C}$ [see \aref{subsec:single-qubitME}], reading
\begin{equation}
    \label{eq:ME}
    \dot{\rho} = -i \left[\Delta \sigma^\dagger \sigma + \Omega \left({\cal C}^\dagger\sigma + \Hc\right), \rho\right] + \gamma(\Delta) {\cal D}_\sigma[\rho]\,,
\end{equation}
where $\mathcal{D}_O[\rho] = O\rho O^\dagger-\{O^\dagger O, \rho\}/2$ is the standard dissipator of the qubit's population, $\Omega = g/ \sqrt{\mathcal{A}}$ with $\mathcal{A}^{-1} = \sqrt{3}{/}\pi{-}1{/}3$, is the qubit-mode interaction strength (proportional to the vacuum Rabi frequency on resonance) and $\gamma(\Delta)$ is the $\Delta$-dependent decay rate into bulk modes (we 
incorporated factor $\sqrt{\cal A}$ in the definition of $\cal C$).
For $|\Delta|\ll J$, we can approximate [see \aref{sec:gamma}]
\begin{equation}
    \gamma(\Delta) \simeq\frac{2g^2|\Delta|}{\sqrt{3}J^2}\,,
\end{equation}
which vanishes for $\Delta\to 0$ consistent with the aforementioned DOS features.
\eq\eqref{eq:ME} defines an effective dissipative Jaynes-Cummings model \cite{lorenzo_quantum_2017} where a lossy qubit coherently couples to a fictitious {\it cavity} embodied by the superposition mode ${\cal C}$.
We point out that the validity of \eqref{eq:ME} relies crucially on whether mode ${\cal C}$ is normalized so as to ensure $[\mathcal{C}, \mathcal{C}^\dagger]=\mathbb{1}$.
To show this, we notice that in real space $\mathcal{C} = \sqrt{\mathcal{A}}\sum_{nm} c(n,m) a_{nm}$ with [\cf \eq\eqref{eq:edge-states}]
\begin{equation}
    \label{eq:cavitymode}
    c(n,m) = \int\limits_{\frac{2\pi}{3}{<}|k|{<}\pi}\!\mathrm{d} k \,\varepsilon^*_k(0,0)\,\varepsilon_k(n,m)\,,
\end{equation} 
and, since $\sum_{n,m} |c(n,m)|^2=\mathcal{A}^{-1}$ in the thermodynamic limit, mode ${\cal C}$ is normalized.
Accordingly, $\mathcal{A}$ is naturally interpreted as a (2D) \textit{cavity volume} yielding a finite Rabi frequency.
This emergent cavity mode however has a non-standard shape, whose spatial distribution is plotted in \ref{fig:fig2a}(a) using \eq\eqref{eq:cavitymode}.
The mode has support on the $a$-sublattice only, which is inherited from edge modes ${\cal E}_k$.
More remarkably,
along the edge direction (resonators $a_{n=0\,m}$), the integral in \eqref{eq:cavitymode} yields $c(0,0) = \mathcal{A}^{-1}$, $c(0,\pm 1) = \sqrt{3}/(4\pi)-1/3$ while for $|m|\ge 2$
\begin{align}\label{eq:c0m}
    c(0,m) = \frac{2}{3}\Big[s(|m|-1)+s(|m|)+s(|m|+1)\Big]\,, && \text{where} \;\;
s(x) = \frac{\sin{(2\pi x/3)}}{2\pi x/3}\,.
\end{align}
\ref{fig:fig2a}(b) shows a plot of $c(0,m)$ versus $m$, which is easily checked to scale as $\sim |m|^{-2}$ for large $|m|$.
Towards the bulk, on the other hand, the mode localization is stronger but still power law since $c(n{\gg} 1,m) \sim |n|^{-2}$ [see \aref{sec:scalingC}].
Thus, the mode shows up an anomalous power-law localization around the qubit with nodes on $b$-resonators, yet remains normalizable even in the thermodynamic limit, yielding finite cavity volume $\mathcal{A}$ and Rabi frequency $\Omega$.
This is a non-obvious characteristic, which \eg does not occur for zero-dimensional defect modes in the bulk \cite{pereira_graphene_2006}, where instead a \textit{quasi-}bound state appears.
We note that the mode’s power-law scaling cannot be explained by standard linear band touchings \cite{kim2025real,rhim_singular_2021} due to the absence of compact localized states \cite{park2024quasi} [see also \aref{app:CLS}].
This scaling originates from the restricted support of edge modes in the BZ, which is topological in nature. Indeed, the limited support of edge modes introduces a non-analyticity at $k=2\pi/3$ in the integrand of Eq. (7), which prevents a full strong localization of the cavity mode. In other words, due to the limited support (which is a consequence of topology), the interference of edge modes constructing the cavity mode is \textit{incomplete}, resulting in a milder scaling at long distances.
We point out that the mechanism behind the emergence of this mode is fundamentally different from conventional bound states in photonic bandgap lattices in which cases a localized defect pulls a state into the bandgap (where photonic propagation is intrinsically forbidden), hence resulting in an exponential localization around the defect.
This scaling is predicted to survive also in presence of moderate next-to-nearest neighbor hoppings, as discussed in \aref{sec:scalingC}.

\begin{figure}[t]
    \centering   
    \includegraphics[width=0.8\textwidth]{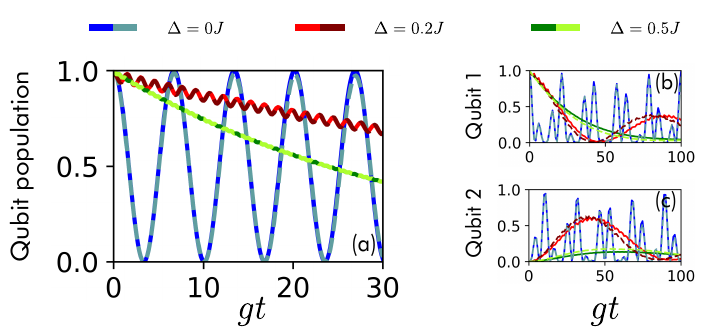}
    \caption{(a) Excited-state population versus time for an initially excited qubit for different values of $\Delta/J$.
(b)-(c) Excitation transfer between two emitters, coupled to $a_{00}$ (qubit 1) and $a_{02}$ (qubit 2), for the initial state $|e_1 g_2\rangle$ and same values of $\Delta$ as in (a).
In (a)-(c), where $g=0.05J$, solid and dashed lines are respectively the analytical solutions using \eqref{eq:ME} and numerical solutions based on the full model \eqref{Htot} for $600{\times}600$ units cells.
The common color legend is reported at the top.}
    \label{fig:fig2c}
\end{figure}

Based on \eqref{eq:ME}, the dynamics of an initially-excited qubit generally undergoes damped vacuum Rabi oscillations with oscillation frequency $\Omega_R(\Delta)=\sqrt{\Delta^2+4\Omega^2}$ and damping rate $\gamma(\Delta)$.
For small $\Delta/J$, the coherent coupling to mode ${\cal C}$ dominates over dissipation into bulk modes, whereas the opposite holds for $\Delta$ large enough.
This is displayed in \ref{fig:fig2c}(a) for $g=0.05 J$ and different values of $\Delta/J$, showing that the solution predicted through \eqref{eq:ME} matches the exact numerical simulations based on the full Hamiltonian \eqref{Htot}.
The effectiveness of \eqref{eq:ME} even for $\Delta=0$ is remarkable, as the model neglects the aforementioned DOS singularity at the Dirac points.
For emitters coupled to the lattice {\it bulk}, a similar singularity triggers non-Markovian effects, most notably fractional decay \cite{gonzalez2018exotic,redondo2021quantum}, due to the presence of the previously-mentioned quasi-bound state, and fast non-Markovian relaxation at intermediate times.
In our system, however, the presence of the FB completely removes such exotic bound state, ensuring that no additional population trapping takes place. This can be shown by computing the residue at $z=i0^+$ of the system's resolvent [see Appendix C.4], which is found to vanish for $\Omega\not= 0$. In other words, the presence of \textit{real} bound states (i.e., the states formed by the qubit dressed with the cavity mode) completely washes out the \textit{quasi-bound} states usually emerging from a Dirac's cone singularity in the local density-of-states \cite{gonzalez2018exotic}.
Furthermore, the FB introduces a power-law tail in the long-time limit $t\gg\Omega^{-1}$  [see \aref{subsec:non-Markov}], thereby allowing for the occurrence of several full Rabi cycles as in \ref{fig:fig2c}(a). In any case, we show in \aref{subsec:non-Markov} that this power-law contribution is proportional to $(g/J)^2$, meaning that it will yield negligible contribution to the dynamics at any time in the weak-coupling regime $g\ll J$.

In presence of next-to-nearest neighbor (NNN) hoppings, the band of edge modes will acquire a finite spectral width, proportional to $J'$ (the NNN hopping rate). Mathematically, this would prevent the definition of the cavity mode as an eigenstate of the bath Hamiltonian. Nonetheless, as long as $g\gg J'$, we can still make use of the effective model (where now $\mathcal{C}^\dagger|\rm vac\rangle$ is an \textit{approximate} eigenstate of the bath Hamiltonian). The same reasoning applies to the many-qubit generalization of our model, discussed in \autoref{sec:many-qubits-generalization}.

\section{Cavity-QED model for $\mu=0$: many qubits}
\label{sec:many-qubits-generalization}

We next generalize to many qubits [\cf\eq\eqref{Htot}] coupled to \textit{distinct} resonators (see \aref{subsec:many-qubit} for details).
Under conditions analogous to those underpinning \eqref{eq:ME} plus the standard assumption that emitters are not too distant so as to make retardation effects negligible \cite{windt_retardation_2025}, the bulk modes can be treated as a common Markovian bath for the qubits.
The interaction of qubit $j$ to modes ${\cal E}_k$'s can again be arranged as the coupling to an effective cavity mode $\tilde{\mathcal{C}}_j$ defined analogously to ${\cal C}$ [\cf\eq\eqref{eq:cavitymode}] but the replacement $c(n,m)\rightarrow c(n,m{-}m_j)$.
Zero-frequency modes $\{\tilde{\mathcal{C}}_j\}$, which are as many as the number of qubits $N_q$, are however non-orthogonal.
Still, similarly to \rref\cite{de_bernardis_light-matter_2021}, they can be orthonormalized yielding the new set of modes 
\begin{equation}
    \mathcal{C}_j = \sum_{j'} (M^{-1})_{jj'} \tilde{\mathcal{C}}_{j'}\,,
\end{equation}
with the $N_q{\times}N_q$ invertible matrix $\mathbf{M}$ such that
\begin{equation}
    \mathbf{M}\mathbf{M}^\dagger = \mathbf{ P}\,,
\end{equation}
where $P_{ij}=c(0,m_{j}-m_j)$.
The resulting modes fulfill $[\mathcal{C}_i, {\cal C}^\dag_j]= \delta_{ij}$  and thus can be viewed as independent cavity modes (one for each qubit).
Note that, although orthogonal, these are still spatially overlapping, hence each emitter $i$ is generally coupled to all modes ${\cal C}_{j}$'s (not only to ${\cal C}_i$).
Based on the above, a generalization of the procedure leading to \eq\eqref{eq:ME} returns the collective master equation 
\begin{equation}
    \label{eq:ME2}
    \dot{\rho} = {-}i \left[\Delta \sum_j \sigma_j^\dagger \sigma_j + g \sum_{ij}({\cal C}_i^\dagger M_{ij} \sigma_j{+} \Hc), \rho\right]+ \mathcal{L}[\rho],
\end{equation}
where ${\cal L}\left[\rho\right] = \sum_{ij} \gamma_{ij}(\Delta) \left[\sigma_j \rho \sigma_i^\dag - \{\sigma_i^\dagger\sigma_j, \rho\}/2\right]$ is a many-qubit dissipator depending on rates $\gamma_{ij}$.
For $\Delta$ small enough, these can be approximated as 
\begin{equation}
    \gamma_{ij}(\Delta) \simeq \frac{2 g^2 |\Delta|}{\sqrt{3}J^2}\cos{\left(\frac{2\pi}{3} |m_i - m_j|\right)}\,,
\end{equation}
vanishing for $\Delta\to 0$. Notice the oscillatory behavior of $\gamma_{ij}=\cos{\left(2\pi |m_i - m_j|/3\right)}$. Unlike the standard case of qubits coupled to a 1D linear waveguide, here the periodicity of this oscillation is not set by the qubits' common frequency, rather by the property of the bath itself, in particular the position of the Dirac cone ($k=2\pi/3$).

\ref{fig:fig2c}(b-c) shows the population dynamics of two qubits coupled to resonators  $a_{00}$ and $a_{02}$ with the former (latter) initially excited (unexcited).
A generally incomplete (due to losses into bulk modes) excitation transfer occurs for $|\Delta|>0$.

For $\Delta=0$ (resonant coupling to edge modes), however, state transfer is achieved with fidelity $0.93$ [see \ref{fig:fig2c}(b-c)], quantified by the maximum value reached by the second qubit's excited-state population\footnote{For the present class of dynamics, perfect excitation transfer directly implies perfect quantum state transfer between the qubits \cite{bose_quantum_2003}.}.
The characteristic beatings, absent in the one-qubit dynamics of \ref{fig:fig2c}(a), are a signature that {\it two} effective modes (${\cal C}_1$ and ${\cal C}_2$) participate in the dynamics.
\begin{figure}[h]
    \centering   
    \includegraphics[scale=1]{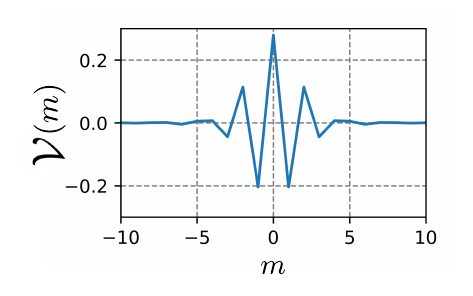}
    \caption{Dispersive qubit-qubit interaction potential $\mathcal{V}(m)$ [\cf\eq\eqref{eq:potential}] for $\mu = 0.3J$ and $\Delta {=} 0.05J$.
Notice the shape similar to $c(0,m)$ i.e., the amplitude of the cavity mode along the edge [c.f.
\ref{fig:fig2a}(b) and \eq \eqref{eq:c0m}].}
    \label{fig:fig2b}
\end{figure}

\section{Dispersive dipole-dipole mediated interactions for $\mu>0$}

In general, photon-mediated qubit-qubit interactions can be made dissipationless, entailing perfect state transfer, in the dispersive regime \cite{douglas_quantum_2015,gonzalez-tudela_subwavelength_2015,bello2019unconventional,Sundaresan2019,scigliuzzo_controlling_2022,zhang2023superconducting,leonforte_quantum_2024,LeonforteVDS}.
The cavity-QED model \eqref{eq:ME2}, however, does not admit a dispersive regime \cite{ritsch_cold_2013} since this would in particular require to detune emitters from ${\cal C}$ by increasing $|\Delta|$, which however comes at the cost of enhanced dissipation into bulk modes (as $\gamma_{ij}\propto |\Delta|$).
This regime is instead possible for $\mu>0$ [\cf\eq\eqref{Htot} and \ref{fig:fig1}(c)], which opens a gap in the bath spectrum while spectrally shifting modes $\mathcal{E}_k$'s by $\mu$.
By tuning the qubits within the gap in a way that $|\Delta|<\mu$, their dynamics becomes dissipationless being described by the effective Hamiltonian
\begin{equation}
    \mathcal{H}_{\rm eff} = \sum_{i,j} \mathcal{K}_{ij}\, \sigma_i^\dagger \sigma_j\,,
\end{equation}
where the interaction potential reads
\begin{equation}
    {\cal K}_{ij}=\frac{g^2}{\Delta-\mu}{\cal V}(m_i-m_j)\,,
\end{equation}
being
\begin{equation}
    \label{eq:potential}
    {\cal V}(m)= c(0,m)+(\Delta{-}\mu) G_{\rm bulk}(m)\,.
\end{equation}
Here, $G_{\rm bulk}(m)$ is the matrix element of the bulk modes Green's function at frequency $\Delta$ between sites ${\R}_i$ and ${\R}_{j}$, which depends only on distance $m=m_i-m_j$ [see \aref{sec:Kij}].
The dipole-dipole interaction potential \eqref{eq:potential} thus comprises a term due to edge modes, having exactly the same shape as mode ${\cal C}$ [\cf\eq\eqref{eq:c0m} and \ref{fig:fig2a}(b)] plus a contribution due to the dispersive coupling to bulk modes.
The shape of ${\cal V}(m)$, plotted in \ref{fig:fig2b}, is very similar to $c(0,m)$
exhibiting an analogous power-law scaling $\sim \!|m|^{-2}$.
Interestingly, as shown in Appendix~I, in the limit $m\gg 1$ the contribution of the dispersive bands to the interaction potential becomes negligible, so that ${\cal V}(m)\to c(0,m)$, in the regime $\mu\gg J$ and $\Delta\simeq 0$. In this regime, the interactions are mediated exclusively by the topologically protected edge modes.

\begin{figure*}[h]
    \centering
    \includegraphics[width=0.8\textwidth]{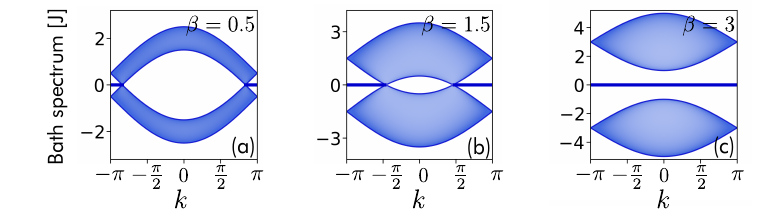}
    \caption{Bath spectrum for different values of the anisotropy parameter $\beta$.}
    \label{fig:fig3}
\end{figure*}

\section{Anisotropic lattice}

\begin{figure*}[h]
    \centering
    \includegraphics[width=0.8\textwidth]{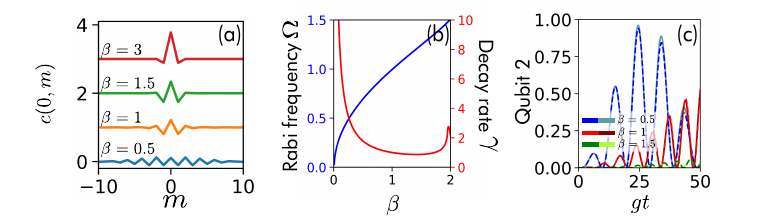}
    \caption{(a) Amplitude $c(0,m)$ for different values of the anisotropy parameter $\beta$.
For clarity, curves for $\beta= 1, 1.5, 3$ are shifted upward by $1, 2, 3$, respectively.
(b) Rabi frequency $\Omega$ (blue) and loss rate $\gamma$ (red) versus $\beta$ for $\Delta=0.05J$ (rescaled to their $\beta=1$ values).
(c) Qubit 2 population for two emitters at distance $m_1-m_2=6$ with qubit 1 (2) initially excited (unexcited) for $\beta= 0.5,1, 1.5$ and $\Delta = 0$.
In (c), we set $g=0.05 J$.}
    \label{fig:fig3b}
\end{figure*}

Much of the above physics stems from the property that modes ${\cal E}_k$ have support only on the  {\it limited} region of the BZ external to the pair of Dirac points [\cf\ref{fig:fig1}(b-c)].
The size of this region can be changed by making the honeycomb lattice {\it anisotropic}, allowing us to study the impact of anisotropy on the previous cavity-QED picture.
We thus relax the constraint
$J_{\langle\R,\R'\rangle}=J$ [\cf\eq\eqref{Htot}] and set now to $\beta J$ the hopping rate between resonators $b_{\mathbf{R}}$ and $a_{\mathbf{R}+{\bf e}_1}$ [see bottom of \ref{fig:fig1}(a)] with $\beta>0$ an anisotropy parameter;
hence the previous system retrieved for $\beta=$1. For the sake of argument, we set $\mu=0$.
Two bands of bulk modes and a set of zero-frequency edge modes ${\cal E}_k$ still occur [see \ref{fig:fig3}(a-c)] \cite{kohmoto2007zero}.
For $\beta < 2$, the modes ${\cal E}_k$ still arise outside the two Dirac points, which however now lie at $k_D = 2\arccos(\beta/2)$ and $-k_D$.
As $\beta$ decreases below 1, the support of the partial flat band progressively shrinks, whereas it broadens as $\beta$ increases above 1 until matching the full BZ for $\beta=2$.
As long as $\beta<2$, one can still define a cavity-QED  model as in \eq\eqref{eq:ME} with the superposition mode ${\cal C}$, $\Omega$ and $\gamma$ now $\beta$-dependent.
Along the edge direction, the amplitude of ${\cal C}$ can be still analytically computed as ($|m|\geq2$)
\begin{align}
    c(0,m)\propto s_\beta(|m|-1)+ (2-\beta^2) s_\beta(|m|)+ s_\beta(|m|+1) \,, && \text{with}\;\;s_\beta(x) = \frac{\sin{(k_Dx)}}{k_Dx}\,.
\end{align}
Compared to \eq\eqref{eq:c0m}, this still scales as $\sim |m|^{-2}$ but, remarkably, has larger width as $\beta$ is reduced below 1, as shown in \ref{fig:fig3b}(a). This can be understood through the same incomplete-interference argument behind the scaling of the cavity mode for $\beta=1$. Indeed, decreasing $\beta$ causes $k_D$ to approach $\pi$, hence shrinking the support of the edge modes on the BZ. As a consequence, this leads to a larger envelope.
Correspondingly, as shown in \ref{fig:fig3b}(b), the Rabi frequency $\Omega\propto {\cal A}^{-1/2}$ decreases, while the single-qubit loss rate, approximated as
\begin{equation}
    \gamma(\Delta) \simeq \frac{2g^2 |\Delta|}{J^2\beta \sqrt{ 4 {-} \beta^2}}\,,
\end{equation}
becomes stronger.
The broadening of ${\cal C}$ substantially impacts excitation transfer in the case of many qubits.
An instance is reported in \ref{fig:fig3b}(c), where state transfer between two qubits separated by six unit cells is achieved with fidelity $\simeq 0.94$ for $\beta=0.5$ in a time $g\tau\simeq 25$.
In contrast, at time $\tau$ the transfer process is still in the early stages for $\beta=1$ and negligible for $\beta=1.5$.
For $\beta>2$, Dirac points no longer exist with
the edge modes now forming a flat band gapped from bulk modes and extended across the full BZ, which results in finite-ranged cavity-QED dynamics.
Nonetheless, even in this case compact states cannot be defined unlike standard FBs \cite{di2025dipole}.

\section{Experimental implementation}

A proof-of-principle experimental demonstration of these phenomena is within reach in circuit-QED platforms using transmon qubits emitting into a lattice of capacitively-coupled LC resonators~\cite{underwood2016imaging, morvan2021observation, kim2021quantum, scigliuzzo_controlling_2022, o2025circuit}.
Typical parameters are $J/2\pi =100-\SI{200}{\MHz}$, $g/2\pi \sim \SI{20}{\MHz}$ and a resonator frequency $\omega_r/2\pi\sim  \SI{6}{\GHz}$.
With $g/2\pi = \SI{20}{\MHz}$, the resulting dynamics occurs on a $\sim \SI{330}{\ns}$ timescale, well within the coherence times of transmon qubits ($\gtrsim \SI{5}{\us}$)~\cite{kim2021quantum,scigliuzzo_controlling_2022}.
The phenomenology remains robust against various experimental non-idealities [see \aref{sec:experiment}], such as e.g.
a parasitic qubit decay rate of $\gamma_q/2\pi\sim \SI{100}{\kHz}$, relative frequency disorder of the bare resonators up to $0.5\% \,\omega_r$, exceeding disorder levels in state-of-the-art implementations \cite{jouanny2025high, ferreira2021collapse}, or finite-size effects, which do not spoil appreciably neither single nor two-qubit oscillations in a system with $\gtrsim 450$ resonators.
On top of that, another source of imperfections comes from next-nearest-neighbor hopping terms, spoiling the \textit{flatness} of the band of edge modes.
In our circuit, these are introduced by adding stray capacitances $C_s$, to the circuit Hamiltonian, coupling bare resonators beyond nearest-neighbors.
Simulations are then performed for values of $C_s$ introducing uniform next-nearest-neighbor hopping up to $\sim 5\% J$, showing the resilience of the effect.
Even considering all of the above together in a single simulation, the effects remain clearly observable, as discussed in \aref{sec:experiment}.
Beyond circuit QED, these phenomena could be observed in platforms such as exciton-polaritons \cite{Chochon:25,jamadi_2021_optically,amo_tilted_2019,bouscal2024systematic} and photonic crystal lattices, which support both graphene-like physics \cite{rechtsman2013topological,barik2018topological} and strong emitter coupling \cite{barik2018topological,arcari2014near-unity}.
\section{Conclusions}
We studied a system of qubits emitting into the boundary of a 2D lattice sustaining dispersionless edge modes defined on a limited region of the BZ (partial flat band).
The essential physics is captured by an effective cavity-QED model, where the emitter leaks into bulk modes and at the same time couples to an emergent mode showing an unusual power-law localization inherited from the scaling behavior of the partial FB \cite{park2024quasi}.
The range of this mode, which is connected to the length of the flat band support in the BZ, affects substantially photon-mediated qubit-qubit interactions, benefiting in particular long-range state transfer.
We point out that coupling qubits to {\it bulk} resonators is experimentally demanding with current circuit-QED technology, which provides a further major motivation/advantage for the present study.
From the perspective of topological photonics \cite{ozawa_topological_2019}, our work introduces a new paradigm of light-matter interactions when emitters are coupled to edge modes of a 2D lattice, where the effective dynamics resemble more the ones of open cavity-QED systems.
This has to be contrasted with other paradigmatic instances of qubits coupled to the edge modes of a 2D topological bath, where the typical effective picture is the one of qubits coupled to a 1D chiral waveguide.
On the other hand, it unveils a novel system yielding power-law photon-mediated interactions, which are known to lead to a number of phenomena such as exotic many-body phases \cite{gong2016topological,gong2016kaleidoscope,vzunkovivc2018dynamical,islam2013emergence,sanchez_limits_2020}.
\section{Acknowledgements}
X.S. acknowledges financial support from China Scholarship Council (Grant No.~202208410355).
We acknowledge financial support from European Union-Next Generation EU through projects: Eurostart 2022 ‘Topological atom-photon interactions for quantum technologies’;
PRIN 2022–PNRR No. P202253RLY
‘Harnessing topological phases for quantum technologies’; THENCE–Partenariato Esteso
NQSTI–PE00000023–Spoke 2 ‘Taming and harnessing decoherence in complex networks’.
P.S.
acknowledges support from the Swiss State Secretariat for Education, Research and Innovation (SERI) under contract numbers UeM019-16 and MB22.00081 / REF-1131 -52105.
P.S. also acknowledges support from the Swiss National Science Foundation (SNSF), Grant number 200021\_200418 / 1.

\appendix

\makeatletter
\let\orig@seccntformat\@seccntformat 
\def\@seccntformat#1{%
  \ifcsname formatapp@#1\endcsname
    \csname formatapp@#1\endcsname
  \else
    \orig@seccntformat{#1}%
  \fi
}
\def\formatapp@section{Appendix \thesection.\quad}
\makeatother

\section{Normal modes of the lattice Hamiltonian $H_B$} \label{app:geometry}

Pristine graphene has the structure of a composite 2D triangular Bravais lattice \cite{Ashcroft:102652}.
A lattice site (resonator) has position given by $\mathbf{R}_{\nu} = n \mathbf{e}_1 + m \mathbf{e}_2 + \mathbf{d}_\nu$, where $(n,m)$ are integers while 
\begin{align}
    \mathbf{e}_1 = a \sqrt{3} \left(\frac{\sqrt{3}}{2},\frac{1}{2}\right)\,, && \mathbf{e}_2 = a\sqrt{3}\,\left(0,1\right)\,, &&\mathbf{d}_\nu = \nu a \left(\frac{1}{2},\frac{\sqrt{3}}{2}\right)
\end{align}
are respectively primitive vectors ($\mathbf{e}_{1(2)}$) and basis vectors ($\mathbf{d}_\nu$) with $\nu=0,1$ the sublattice index, and where $a$ is distance between nearest-neighbor resonators defined as $a = |\mathbf{R}_{1}-\mathbf{R}_{0}|$.
This entails e.g. that $|\mathbf{e}_1| = a\sqrt{3}$, corresponding to the distance between unit cells located at $(n,m)$ and $(n\pm1,m)$ (same goes for $|\mathbf{e}_2|$).
Sublattice $\nu{=}0$ ($\nu{=}1$) corresponds to the $a$- ($b$-) sublattice in the main text.
\noindent
The reciprocal lattice has primitive vectors
\begin{align}
    \mathbf{b}_1 = 2\pi \frac{\mathbf{Q}.\mathbf{e}_2}{\mathbf{e}_1 \cdot \mathbf{Q}\mathbf{e}_2} = \frac{2\pi}{V} a\sqrt{3}\left(1,0\right) \,, &&     
    \mathbf{b}_2 =  2\pi \frac{\mathbf{Q}\mathbf{e}_1}{\mathbf{e}_2 \cdot \mathbf{Q}\mathbf{e}_1}= \frac{2\pi}{V} a\sqrt{3} \left(-\frac{1}{2},\frac{\sqrt{3}}{2}\right)\,,
\end{align}
where $V = |\mathbf{e}_1 \cdot \mathbf{Q}\mathbf{e}_2| =  a^2 3\sqrt{3}/2$ is the volume of the primitive unit cell and $\mathbf{Q}$ is the $2\times 2$ matrix describing a counter-clockwise $\frac{\pi}{2}$ rotation on the $xy$ plane.
In the following, we will set $a=1$ and assume $N_1$ ($N_2$) unit cells along the $\mathbf{e}_1$ ($\mathbf{e}_2$) direction.
Based on the above, the Hamiltonian $H_B$ [see \eq \eqref{eq:HB} in the main text]  with a zigzag edge embodied by all cells having $n=0$ can be conveniently arranged as
\begin{equation}
    \label{eq:app-zigzag}
    H_B = \sum_m h_B(m)\,
\end{equation}
with
\begin{equation}
    h_B(m) = \mu \sum_{n=0}^{N_1-1} \left(a^\dagger_{nm} a_{nm} - b^\dagger_{nm} b_{nm}\right) +J \sum_{n=1}^{N_1-1} \left(\beta \,a_{n+1 m} + a_{nm}\right)b_{nm}^\dagger +J\sum_{n=0}^{N_1-1} a_{n m+1} b_{nm}^\dagger +\Hc\,,
\end{equation}
which holds even in the case of an anisotropic lattice specified by the parameter $\beta$.
The thermodynamic limit discussed in the main text is obtained for $N_1,N_2\to\infty$.
We next review how to map the lattice Hamiltonian $H_B$ into a set of uncoupled 1D Rice-Mele models \cite{asboth2016short}, which dramatically simplifies the calculation of its normal modes and spectrum.
\subsection{Mapping to uncoupled Rice-Mele models}\label{app:SSH-der}

While the edge enforces a hard-wall boundary condition along the direction $\mathbf{e}_1$, the direction $\mathbf{e}_2$ (along the edge) has no boundaries and we can accordingly enforce periodic boundary conditions (BCs).
Along the latter direction, the lattice is therefore translationally invariant allowing to take advantage of the Bloch theorem \cite{bloch_ber_1929}.
Accordingly, we perform a partial Fourier-transform of real-space bosonic ladder operators defined by
\begin{align}
    \label{eq:partial-fourier}
    a_{nm} =\frac{1}{\sqrt{2\pi}}\sum_{k}e^{ik\left[m+\phi_{a}(n)\right]}\,a_{kn}\,,&& 
    b_{nm} =\frac{1}{\sqrt{2\pi}}\sum_{k}e^{ik\left[m+\phi_{b}(n)\right]}\,b_{kn}\,,
\end{align}
where $\phi_{a}(n) = n/2$ and $\phi_{b}(n) = (n+1)/2$ are $n$-dependent phase factors, which are introduced so as to ensure that the transformed Hamiltonian features only real matrix elements.
Operators $a_{kn}$ and $b_{kn}$ fulfill bosonic commutation rules. Substituting this expression for the cavity ladder operators in \eq \eqref{eq:app-zigzag}, after some lengthy algebra, the bath Hamiltonian can be decomposed as $H_B = \sum_k \mathcal{H}_k$, where
\begin{equation}
    \mathcal{H}_k = \mu\sum_{n=0}^{\infty} \left(a_{kn}^\dagger a_{kn}-b_{kn}^\dagger b_{kn}\right) +J \sum_{n=0}^{\infty} \left[\beta\,a_{kn+1} b_{kn}^\dagger + 2\cos{\left(\tfrac{k}{2}\right)}a_{kn}b_{kn}^\dagger + \Hc\right]\,.
\end{equation}
Thus, the model is mapped into a collection of uncoupled 1D Rice-Mele models: each is labeled by $k$ and features an inter-cell hopping rate $\beta J$ and a $k-$dependent intra-cell hopping rate $\Jt = 2J \cos{\frac{k}{2}}$.
The diagonalization of $H_B$ thereby reduces to the diagonalization of $\mathcal{H}_k $, which is carried out next first for bulk modes and then for edge modes.
\subsection{Bulk modes}

Under periodic BCs, the bulk spectrum of $\mathcal{H}_k$ is worked out as
\begin{equation}
    \omega_{\pm}(k,q) = \pm \sqrt{\mu^2 +(\beta J)^2+\Jt^2 + 2\beta J\Jt\cos{q}}=\pm J \,\omega(k,q)\label{wpm}
\end{equation}
where $0\leq q\leq 2\pi$.
The normal mode corresponding to frequency $\omega_\pm(k,q)$ reads
\begin{equation}
    { B}_\pm (k, q) =\frac{\mathcal{N}_\pm(k,q)}{\sqrt{2\pi}} \sum_{n} \Bigg[ \left(2\cos{\frac{k}{2}} + \beta e^{-iq}\right) e^{-iqn} a_{kn} \pm (\omega(k,q)\mp(\mu/J)) \,e^{-iqn} b_{kn}\Bigg]
\end{equation}
where $\left[\mathcal{N}_\pm(k,q)\right]^{-2} = 2\omega(k,q)\left[\omega(k,q)\mp(\mu/J)\right]$ is a normalization constant.
Exploiting the definition of the $a_{kn},b_{kn}$ operators given in \eq \eqref{eq:partial-fourier}, ${ B}_\pm (k, q)$ can be expressed in real space as
\begin{equation}
    \begin{split}
    { B}_\pm (k, q) = \frac{\mathcal{N}_\pm(k,q)}{2\pi} \sum_{n,m} &\Bigg[\left(2\cos{\frac{k}{2}} + \beta e^{-iq}\right) e^{-ik(m+n/2)} e^{-iqn} a_{nm}\\
    &\pm (\omega(k,q)\mp(\mu/J)) e^{-ik(m+n/2+1/2)} e^{-iqn} b_{nm}\Bigg]\,.
\end{split}
\end{equation}
The above bulk modes are valid under {\it periodic} BCs.
For \textit{open} BCs, which are those of concern here, the bulk spectrum of ${\cal H}_k$ in \eq\eqref{wpm} still holds while normal modes can be constructed from those under PBCs as the superposition ${\cal B}_\pm (k, q) = \left[{ B}_\pm (k, 2\pi -q)-{B}_\pm (k,q)\right]/(\sqrt{2}i)$ where now $0\leq q\leq\pi$.
This yields
\begin{equation}
    \begin{split}
    {\cal B}_\pm (k, q) = \mathcal{N}_\pm(k,q)\sqrt{\frac{2}{\pi}} \sum_{n=0}^{\infty} &\Bigg[\left(2\cos{\frac{k}{2}}\sin{q(n+1)} +\beta\sin{qn}\right) a_{kn}\\
    &\pm (\omega(k,q)\mp(\mu/J)) \sin{q(n+1)}b_{kn}\Bigg]\,,
    \end{split}
\end{equation}
which can be again expressed in real space using \eq \eqref{eq:partial-fourier}.
\subsection{Edge modes}

In the gapless case i.e., for $\mu{=}0$, Hamiltonian ${\cal H}_k$ reduces to a $k$-dependent SSH Hamiltonian \cite{asboth2016short}.
Accordingly, for $|\Jt|<\beta|J|$, which in the present context corresponds to the condition
\begin{equation}
    \label{eq:support}
    k_D=2\arccos{\tfrac{\beta}{2}}< |k| \leq\pi
\end{equation} 
zero-frequency edge modes appear under open BCs.
Notably, $\pm k_D$ are the values in which Dirac points appear in the bulk modes, meaning that the edge modes will cover the region of the FBZ outside those points.
The edge modes lying on the left edge read \cite{tan_edge_2021} (recall that our lattice is semi-infinite)
\begin{equation}
    \mathcal{E}_k = \mathcal{N}_k \sum_n (-1)^n e^{-n/\lambda_k} a_{kn}\,
\end{equation}
with $\lambda_k^{-1} = \ln\left|{\beta J/\Jt}\right|$ and $\mathcal{N}_k = \sqrt{-\left[2/\beta^2-1+(2/\beta^2)\cos{k}\right]}$.
In real space, edge modes $\mathcal{E}_k$ read
\begin{equation}
    \label{app-edgemodes}
    \mathcal{E}_k = \frac{\mathcal{N}_k}{\sqrt{2\pi}} \sum_{n=0}^\infty\sum_{m} (-1)^n e^{-ik(m+n/2)}e^{-n/\lambda_k} a_{nm}\,,
\end{equation}
showing that $\lambda_k$ embodies a $k$-dependent penetration length.
As expected, those modes have a non-zero envelope only on the $a$-sublattice.
In the gapped case i.e., for $\mu\not =0$, edge modes will also appear, having the same expression in real space given by \eqref{app-edgemodes}.
The only difference is their energy, now equal to $\mu$ i.e., the value of the on-site energy of cavities in the $a$-sublattice.
\section{Properties of the flat band of edge modes}\label{app:CLS}

The frequency-degenerate edge modes ${\cal E}_k$ represent a {\it flat band} (FB) of frequency $\mu$.
A major property of standard FBs is that the corresponding subspace can be spanned through a basis of compact localized modes i.e., modes strictly localized in a finite region of space \cite{rhim_classification_2019}.
This property relies crucially on two assumptions, namely the fact that the FB has support on the full BZ and it is topologically trivial \cite{chen_impossibility_2014}.
Our edge modes ${\cal E}_k$, however, form a {\it partial} FB since their support \eqref{eq:support} is {\it shorter} than the BZ for $\beta <2$.
In general, as shown in Ref. \cite{park2024quasi}, this is enough to prevent the construction of a properly-defined set of compact states spanning the FB subspace.
On the other hand, when $\beta>2$ the edge modes form a \textit{full} FB spanning the whole BZ, which is gapped from the rest of the spectrum.
Thus, one might wonder whether in this case it is instead possible to span the FB subspace using a basis of compact states.
We show next that, unlike standard full FBs, this is not possible in the system under study.
Assuming $\beta \geq 2$, we can follow a general procedure (see \eg \rref \cite{di2025dipole} and references therein) which allows to construct a (non--normalized) compact mode $\phi_{m_0}$ localized around resonator $m=m_0$ along the zigzag edge ($n=0$) as
\begin{equation}
    \phi_{m_0} = \frac{1}{\sqrt{2\pi}} \int_{-\pi}^\pi \mathrm{d} k \sqrt{F(k)} e^{ikm_0} \mathcal{E}_k\,,
\end{equation}
where function $F(k)$ is required to fulfill $F(k) = |\mathcal{N}_k|^{-2}$ with ${\cal N}_k$ the normalization constant appearing in the definition of the edge modes $\mathcal{E}_k$ [see \eq \eqref{app-edgemodes}].
In real space, using \eq \eqref{app-edgemodes}, $\phi_{m_0}$ reads
\begin{align}
    \phi_{m_0} = \sum_{n,m} \frac{(-1)^n}{\beta^{n+1}} \varphi(n,m-m_0)\, a_{nm} &&{\rm with}\,\,\,\,\varphi(n,m) = \frac{1}{2\pi} \int_{-\pi}^\pi \mathrm{d} k \left(2\cos{\frac{k}{2}}\right)^n e^{-ik(n+m)}\,.
\end{align}
This last expression is in fact the Fourier transform of $(\Jt/J)^n$, which can be computed easily with the help of the following identities
\begin{equation}
    \cos^n{(\theta)} = 
    \begin{cases}
        \frac{1}{2^{n-1}} \sum_{r=0}^{r<2n} \binom{n}{2r} \cos{\left[(2r-n)\theta\right]}\,, & \text{odd }n\\
        \frac{1}{2^n} \binom{n}{n/2} + \frac{1}{2^{n-1}} \sum_{r=0}^{r<2n} \binom{n}{2r} \cos{\left[(2r-n)\theta\right]}\,, & \text{even }n\\
    \end{cases}
    \,.
\end{equation}
Using these, function $\varphi(n,m)$ can be expressed as
\begin{equation}
    \label{eq:varphi}
    \varphi(n,m) = 
    \begin{cases}
        \frac{1}{2} \sum_{r=0}^{2r<n} \binom{n}{2r} \delta_{r+2m,0}\,, & \text{odd }n\\
        \binom{n}{n/2}\delta_{n+m,0}+ \frac{1}{2} \sum_{r=0}^{2r<n} \binom{n}{2r} \delta_{r+2m,0}\,, & \text{even }n\\
    \end{cases}
    \,.
\end{equation}
This shows that state $\phi_{m_0}$ cannot be compact since, for any fixed $n$, its amplitude is non zero on ${\sim} n$ resonators.
Furthermore, we can compute the squared-norm of this mode as
\begin{equation}
    \label{eq:squarednorm}
    \begin{split}
        \langle {\rm vac}| \,\phi_{m_0} \phi_{m_0}^\dagger| {\rm vac}\rangle &= \sum_{n=0}^{+\infty} \frac{1}{\beta^{2(n+1)}}\sum_m |\phi(n,m-m_0)|^2 = \frac{1}{\beta^2}\sum_{n=0}^{+\infty}\binom{2n}{n} \frac{1}{\beta^{2n}}\\
        &=\frac{1}{\beta\sqrt{\beta^2-4}}\,,
    \end{split}
\end{equation}
which is finite for $\beta>2$, while diverging at $\beta=2$. These facts together, namely Eqs. \eqref{eq:varphi} and \eqref{eq:squarednorm}, show that if $\phi_{m_0}$ is \textit{not} compact for any values of $\beta \geq 2$, while being normalizable in the thermodynamic limit only for $\beta>2$.
Thus, it is not possible in this case, even when the FB has support on the full BZ and is gapped from the bulk spectrum (if $\beta >2$), to span the FB subspace through a basis of compact localized modes.
At $\beta=2$, this impossibility is a consequence of topology, known as \textit{Wannier obstruction}, which prevents to construct localized Wannier functions using eigenmodes of topologically non-trivial bands \cite{po_fragile_2018, park2024quasi}. We see that this is the case here as, upon inspection of \eq \eqref{eq:support}, even for $\beta=2$ the FB is not \textit{full}, lacking the mode at $k=0$. On the other hand, having $\beta>2$ opens a gap between the now-full FB and the rest of the spectrum, lifting the obstruction. This can be seen by looking at the scaling of $\phi_{m_0}$ at large $n$, which yields
\begin{equation}
    \left|\langle{\rm vac}|a_{nm}\phi_{m_0}^\dagger |{\rm vac}\rangle\right| \simeq \frac{1}{\beta} \left( \frac{2}{\beta} \right)^n \sqrt{\frac{2}{\pi n}} \exp\left( - \frac{2(m - m_0 + n)^2}{n} \right)\,,
\end{equation}
which decreases exponentially for increasing $n$. Thus, in this case, the origin of the lack of compact localization is simply the specific geometry of the lattice under study i.e., the zigzag shape of the boundary.
\noindent
For technical convenience, in the remainder we will often make use of a single-photon Dirac formalism such that, in particular, $|a_{n,m}\rangle=a_{nm}^\dag\!|{\rm vac}\rangle$ (with $|{\rm vac}\rangle$ the field's vacuum) denotes the Fock state where resonator $a_{n,m}$ is populated by a single photon.
Likewise, $|{\cal E}_k\rangle={\cal E}_k^\dag|{\rm vac}\rangle$ and $|{\cal B}_\pm^\dag (k,q)\rangle={\cal B}_\pm^\dag (k,q)|{\rm vac}\rangle$ are single-photon edge and bulk states, respectively.
\section{Derivation of the effective cavity-QED model}

\subsection{Markovian master equation: brief review}

Let an open system $S$ be in contact with a bath $B$ with an interaction Hamiltonian of the form $H_{\rm SB}(t) = \sum_i X_i(t) \otimes E_i(t)$ (in the interaction picture), where $X_i$ ($E_i$) are operators on $S$ ($B$).
Introducing operators $\Pi(\varepsilon)$ each of which projects $S$ onto the eigenspace of energy $\varepsilon$, one can expand $X_i$ as $X_i(\omega) = \sum\limits_{\varepsilon'-\varepsilon=\omega} \Pi(\varepsilon) X_i \Pi(\varepsilon')$.
Hence, the interaction Hamiltonian can be expressed as
\begin{equation}\label{eq:HSBt}
    H_{\rm SB}(t) = \sum_{i,\omega} e^{-i\omega t} X_i(\omega) \otimes E_i(t) = \sum_{i,\omega} e^{i\omega t} X_i^\dagger(\omega) \otimes E_i^\dagger(t)\,.
\end{equation}
Next, starting from the Redfield master equation and performing the Markov and weak-coupling approximations, one obtains the integro-differential equation for the reduced density matrix of $S$ \cite{breuer2002theory}
\begin{equation}
    \frac{\mathrm{d}}{\mathrm{d}t}\rho_S = -\int_0^\infty \mathrm{d} s \mathrm{Tr}_B\left\{[H_{\rm SB}(t), [H_{\rm SB}(t-s), \rho_S(t)\otimes\rho_B]]\right\}\,,
\end{equation}
where ${\rm Tr}_B$ stands for the partial trace over $B$.
Substituting \eqref{eq:HSBt} and keeping only slowly-rotating terms corresponding to nearly-resonant bath modes, one ends up with the master equation
\begin{equation}\label{eq:ME-gen}
    \frac{\mathrm{d}}{\mathrm{d}t} \rho_S(t) = \sum_{ij} \sum_\omega \Gamma_{ij}(\omega) \left[X_j(\omega)\rho_S(t) X_i^\dagger(\omega) - X_i^\dagger(\omega) X_j(\omega) \rho_S(t)\right] + \hc\,,
\end{equation}
where $\omega$ is the characteristic frequency of $S$ while matrix $\Gamma_{ij}(\omega)$ is given by 
\begin{equation}\label{eq:gamma}
    \Gamma_{ij}(\omega) = \int_0^\infty \mathrm{d} s\, e^{i\omega s} \mathrm{Tr}_B\left\{E_i^\dagger(t)E_j(t-s)\rho_B\right\} \equiv \frac{1}{2}\gamma_{ij}(\omega) + i\, S_{ij} (\omega)\,.
\end{equation}
The real part of this matrix describe the irreversible decay rate into the bath, while the imaginary part renormalizes the free Hamiltonian of $S$.
\subsection{Single-qubit master equation}
\label{subsec:single-qubitME}

The Hamiltonian of the total system can be rearranged as
\begin{equation}
    H = H_{\rm JC} +\Delta \sigma^\dagger \sigma + \sum_{k,q,\nu} \omega_\nu (k,q)\, \mathcal{B}^\dagger_\nu(k,q)\mathcal{B}_\nu(k,q) + g \sum_{k,q}\sum_{\nu=\pm} \left( f_\nu(k,q) \mathcal{B}^\dagger_\nu(k,q)\,\sigma +\hc\right)\,,
\end{equation}
where
\begin{equation}
    H_{\rm JC} = g \sum_{|k|>k_D} \left( \frac{\mathcal{N}_k}{\sqrt{2\pi}} \mathcal{E}_k \sigma^\dagger + \hc\right)
\end{equation}
is the coupling Hamiltonian describing the interaction between the qubit and all edge modes $\{{\cal E}_k\}$.
Here, $k_D=2\arccos{(\beta/2)}$ and $\varepsilon_k(0,0) = \mathcal{N}_k/\sqrt{2\pi}$ is the amplitude of the edge mode $\mathcal{E}_k$ at site $a_{00}$ to which the atom is coupled.
Also, $f_\pm(k,q) = \langle a_{00} | \mathcal{B}_\pm(k,q) \rangle$ is the amplitude of the bulk mode $\mathcal{B}_\pm(k,q)$ at the resonator coupled to the emitter, reading
\begin{equation}
    f_\pm(k,q) = \frac{1}{\pi}\mathcal{N}_\pm(k,q)\left[ 2\cos{\frac{k}{2}}\sin{q}\right] = \frac{\sqrt{2}}{\pi}\frac{\cos{\frac{k}{2}}\sin{q}}{\left[\omega(k,q)\right]^2}
\end{equation}
in the $\mu{=}0$ case.
Since $f$ turns out to be independent of the band index $\nu$, we will drop such index in the rest of this subsection.
The Hamiltonian $H_{\rm JC}$ can be arranged as [\cf\eq(5) in the main text]
\begin{equation}
    \label{eq:JC-app}
    H_{\rm JC} = g\left[ \sigma^\dagger \left(\sum_{|k|>k_D} \frac{\mathcal{N}_k}{\sqrt{2\pi}}  \mathcal{E}_k + \hc\right)\right] \equiv \Omega \left(\sigma^\dagger \mathcal{C} + \hc\right)\,,
\end{equation}
where $\Omega = g/\sqrt{\mathcal{A}}$ and we have introduced a \textit{superposition mode} $\mathcal{C}$ [\cf\eq(6) therein] such that
\begin{equation}
    \mathcal{C} = \sqrt{\mathcal{A}} \sum_{|k|>k_D} \frac{\mathcal{N}_k}{\sqrt{2\pi}}  \mathcal{E}_k = \sqrt{\mathcal{A}} \sum_{nm} c(n,m) a_{nm}\,.
\end{equation}
In this expression, $c(n,m)$ is the amplitude of the mode in real space (which will be computed in a particular case in \autoref{app:Gij}) and $\mathcal{A}$ plays the role of a normalization constant, which can be computed by requiring that $[\mathcal{C}, \mathcal{C}^\dagger] = \mathbb{1}$ i.e., that the mode obeys bosonic commutation rules.
From this, one sees that
\begin{equation}
    \label{eq:commutator}
    [\mathcal{C}, \mathcal{C}^\dagger] = \mathcal{A} \frac{1}{2\pi}\sum_{|k|>k_D} |\mathcal{N}_k|^2 \to \mathcal{A} \frac{1}{\beta^2 \pi} \int_{k_D}^\pi \mathrm{d} k     \left(\beta^2 - 2 - 2\cos k\right)\,,
\end{equation}
which always converges (regardless of $\beta$) in the thermodynamic limit (the precise value is discussed in \autoref{app:Gij}).
We stress that, due to the normalization of the emergent mode ${\cal C}$, the coupling strength to this mode remains finite in the thermodynamic limit.
This takes care of the interaction with the edge modes.
Regarding the bulk modes, these can be treated as an effective bosonic reservoir to which the qubit is coupled to within the Born-Markov approximation.
This can be justified by looking at the local density-of-states of such modes on the resonator $a_{00}$.
Indeed, recall that we are interested in frequencies nearly-resonant with the FB at $\Delta = 0$, which on the other hand is resonant with the Dirac point emerging from the bulk modes dispersion law [see \eq \eqref{wpm}].
Around this point, it is well known \cite{Girvin_Yang_2019} that the density-of-states of graphene bulk modes scale linearly with their energy, being exactly zero on resonance with the Dirac point i.e., at $\Delta = 0$.
This means that, since light-matter coupling is indeed proportional to said density, the interaction with bulk modes will be much weaker than the one with the FB of edge modes.
Having said that, replacing in \eq \eqref{eq:gamma} $\omega=\Delta$, $X_1 = X_2^\dagger = \sigma$ and $E_1 = E_2^\dagger = g\sum\limits_{k,q,\nu} f_\nu(k,q)$
yields
\begin{equation}
    \Gamma_{ij}(\Delta) = \delta_{ij}\delta_{i1}\, g^2 \sum_{k,q}\sum_{\nu=\pm} |f_\nu(k,q)|^2 \int_0^\infty \mathrm{d} s \,e^{i(\Delta-\omega_\nu(k,q)) s}\,,
\end{equation}
hence 
\begin{equation}
    \label{eq:diss-single}
    \gamma(\Delta) = 2\pi g^2 \sum_{k,q} \sum_{\nu=\pm} |f(k,q)|^2 \, \delta(\Delta-\omega_\nu(k,q))\,.
\end{equation}
Neglecting the Lamb shift and going back to the Schr\"odinger picture, this leads to master equation (5) in the main text.

We stress that in our model only the interaction with bulk modes is treated within the Born-Markov approximation, while the interaction with the FB (entering the coherent part of the Lindblad superoperator) is treated exactly.

\subsection{Many-qubit master equation}
\label{subsec:many-qubit}

The derivation described in the last section can be generalized to the case of $N_q$ qubits as follows. First, we consider the case in which all qubits are coupled to the same resonator of the \textit{a}-sublattice. In this case, the Hamiltonian describing the interaction 
of the qubits with the FB can be rearranged as
\begin{equation}
    H_{\rm TC} = \Omega \left[\left(\sum_{n=1}^{N_q}\sigma_n^\dag\right)\mathcal{C}+\hc\right]\,,
\end{equation}
where $\Omega$ is the Rabi frequency introduced earlier and the \textit{superposition mode} $\mathcal{C}$ is defined as in the single--qubit case.
Thus, we see that in this case the effective Hamiltonian reduces to a Tavis-Cummings model \cite{tavis_exact_1968}, where a set of identical two-level emitters is coupled to the same cavity mode.
Upon tracing off the bulk modes, the master equation in the Schr\"odinger picture reads
\begin{equation}
    \frac{\mathrm{d}}{\mathrm{d}t}\rho_S =-i[\Delta \sum_{i} \sigma_i^\dagger \sigma_i + H_{\rm TC}, \rho_S] +\gamma(\Delta) \sum_{ij} \left(\sigma_j\rho_S \sigma_i^\dagger -\frac{1}{2}\{\sigma_i^\dagger \sigma_j, \rho_S\}\right)\,,
\end{equation}
which describes an open Tavis-Cummings model with collective dissipation at rate $\gamma(\Delta)$, same as the single-qubit case [see \eq \eqref{eq:diss-single}].

Second, we consider the case in which all qubits are coupled to \textit{different} resonators that is, any two distinct qubits are coupled to distinct resonators belonging to the \textit{a}-sublattice. Under this assumption, the qubits-FB interaction Hamiltonian can be rearranged as
\begin{equation}\label{eq:JC-many}
    H_{\rm MJC} = g \sum_{j} \left(\sigma_j \Tilde{\mathcal{C}}^\dagger_j + 
\hc\right)\,,
\end{equation}
where each ladder operator $\Tilde{\mathcal{C}}_j$ (one for each qubit $j$) has the same expression as ${\cal C}$ (apart from the normalization factor) in the single-qubit case with $m\rightarrow m-m_j$.
These $N_q$ modes, however,  overlap one another and, most importantly, they are not orthogonal. As such, $[\Tilde{\mathcal{C}}_i, \Tilde{\mathcal{C}}^\dagger_j]\neq \delta_{ij}$.
To end up with a well-defined set of $N_q$ modes fulfilling bosonic commutation rules, following \rref \cite{de_bernardis_light-matter_2021} we perform a transformation defined by the $N_q\times N_q$ invertible matrix $\mathbf{M}$ such that $\mathbf{M}\mathbf{M}^\dagger = \mathbf{P}$, where $\mathbf{P}$ is the same as in \autoref{app:Gij}.
Modes $\{\Tilde{\mathcal{C}}_j\}$ are transformed into
\begin{equation}
    \mathcal{C}_i = \sum_j (M^{-1})_{ij}\, \Tilde{\mathcal{C}}_j\,,
\end{equation}
which now fulfill $[{\cal C}_i, {\cal C}^\dagger_j]= \delta_{ij}$.
The 
interaction Hamiltonian \eqref{eq:JC-many} in terms of the new commuting modes reads
\begin{equation}
    H_{\rm MJC} = g \sum_{ij} \left(\sigma_i M_{ij} \mathcal{C}^\dagger_i + \hc\right)\,.
\end{equation}
On the other hand, the bulk modes can be traced out using the same procedure as before.
Notice that in this case, though, each $f(k,q)$ acquires a $k$-dependent phase, depending on the atom position $m_i$.
Proceeding as before, the master equation in Schr\"odinger picture reads
\begin{equation}
    \frac{\mathrm{d}}{\mathrm{d}t}\rho_S =-i[\Delta \sum_{i} \sigma_i^\dagger \sigma_i + H_{\rm MJC}, \rho_S] +\sum_{ij} \gamma_{ij}(\Delta) \left[\sigma_j\rho_S \sigma_i^\dagger -\frac{1}{2}\{\sigma_i^\dagger \sigma_j, \rho_S\}\right]\,,\label{ME-SM}
\end{equation}
where the matrix of decay rates $\gamma_{ij}$ reads
\begin{equation}
    \gamma_{ij}(\Delta) = 2\pi g^2 \sum_{k,q,\nu} |f_\nu(k,q)|^2 \cos{(k|m_i-m_j|)}\, \delta(\Delta-\omega_\nu(k,q))\,.
\end{equation}
The diagonal elements of this matrix coincide with the single-qubit decay rate in the previous section, while the off-diagonal ones feature an additional distance-dependent factor $\cos{k|m_i-m_j|}$. Notice that, as is usual in the waveguide-QED literature, this Markovian master equation holds as long as the qubits are close enough to neglect time-delay effects. We stress that this 
assumption applies to the interaction with the bulk modes (described within Born-Markov approximation), while the interaction with the FB is treated exactly.
Master equation \eqref{ME-SM} describes a set of $N_q$ qubits each being coherently coupled to a set of $N_q$ lossless, non-interacting and frequency-degenerate modes $\{{\cal C}_i\}$.
Additionally, the qubits are in contact with a common bath embodied by the bulk modes.
\subsection{Validity of the Markovian approximation}
\label{subsec:non-Markov}

Rigorously speaking, our previous derivation of the master equation did not take into account that the qubit's self-energy is not analytical at $\Delta=0$, which would in principle invalidate the Markov approximation when the qubit is strictly tuned at $\Delta=0$.
A similar issue was extensively discussed for qubits coupled to the bulk of a honeycomb lattice in \rref \cite{gonzalez2018exotic}.
We show next that, as long as the qubit-photon coupling strength is sufficiently weak, which is the case considered in this work, this non-analyticity introduces only a modest effect in the long-time dynamics.
For the sake of argument, we  consider the case of a single qubit coupled to an edge resonator.
In the single-excitation sector, the total state of the system at time $t$ is written as $|\Psi(t)\rangle = c_e(t) |e\rangle + c_{\mathcal{C}}(t) |\mathcal{C}\rangle + \sum\limits_{k,q,\mu} c_{k,q,\mu}(t) | \mathcal{B}_\mu(k,q)\rangle$.
Assuming $c_e(0)=1$ as the initial condition, the amplitude at time $t$ is determined by
\begin{equation}\label{cet-eq}
    c_e(t) = -\frac{1}{2\pi i}\int_{-\infty}^\infty \mathrm{d} E\, e^{-iEt} G_e(E+i0^+)\equiv -\frac{1}{2\pi i}\int_{-\infty}^\infty \mathrm{d} E \frac{e^{-iEt}}{E-\Delta-\Sigma_e(E+i0^+)}\,.
\end{equation}
This is the inverse Fourier transform of the so-called resolvent or Green function of the system \cite{lambropoulos_fundamental_2000}, defined as $G_e(E) = [E-\Delta-\Sigma_e(E+i0^+)]^{-1}$ (strictly speaking the expectation value of the resolvent for the state $|e, {\rm vac}\rangle$), where the key quantity $\Sigma_e(z)$ is the so-called \textit{self-energy} $\Sigma_e(z)$.
In the present system, this is given by
\begin{equation}
   \Sigma_e(z) = \Sigma_{\text{FB}}(z) + \Sigma_{\text{bulk}}(z)\,,
\end{equation}
where the FB's and bulk's contribution and respectively given by $\Sigma_{\text{FB}}(z)=\frac{\Omega^2}{z}$ and
\begin{align}
\Sigma_{\text{bulk}} (z)= \frac{g^2}{N_1(N_2+1)} \sum_{k,q} \frac{\sin^2(q)\left(2J\cos(k/2)\right)^2}{|\omega_+(k,q)|^2}\frac{z}{z^2-|\omega_+(k,q)|^2}\,.
\end{align}
From now on, we will set $\Delta=0$. The self-energy is non-analytic at $z=0$.
This singularity is similar - though not identical - to the one arising for 
a qubit coupled to the lattice {\it bulk} \cite{gonzalez2018exotic}, in which case it was shown to trigger two kinds of effects: (i) occurrence of fractional decay due to the presence of a (quasi-)bound state (BS) for a lattice of {\it finite} size (finite-size effect) and (ii) occurrence of non-exponential decay $\sim 1/\log^2(t)$ at long times in an {\it infinite} lattice.
The latter, in particular, is due to a branch cut in the self-energy on the negative imaginary semi-axis, yielding a long-time scaling of the qubit's population  $P_e\sim 1/\log^2(t)$.
We next investigate the occurrence of effects (i) and (ii) in our system.
\subsubsection{Fractional decay}

\begin{figure}
    \centering
    \includegraphics[width=0.55\linewidth]{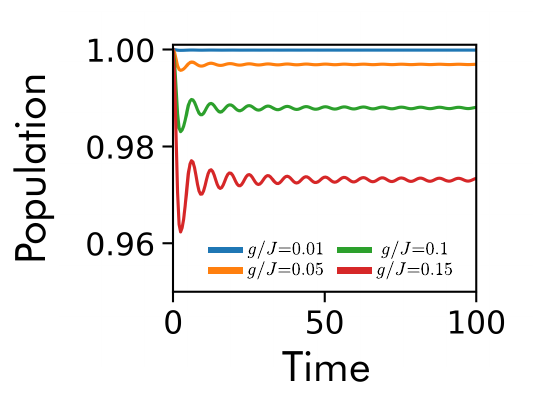}
    \caption{Fractional decay of an initially-excited qubit coupled to resonator $b_{00}$ for different coupling strengths $g/J$ (we set $\beta=1$).
Time is in units of $J^{-1}$.}
    \label{fig:supp_NM}
\end{figure}

Fractional decay occurs when the resolvent $G_e(z)$ has a simple pole at the origin $z=0$ and a corresponding non-zero residue $R_0$.
In the present system, $R_0$ however vanishes as
\begin{equation}
    R_0 = \left. \frac{1}{1-\partial_z \Sigma_{\rm FB}(z)-\partial_z \Sigma_{\rm bulk}(z)} \right|_{z=i0^+} = \lim_{\eta\to0^+} \frac{\eta^2}{\eta^2\left[1+(g/J)^2 g(N)\right]-\Omega^2} = 0\,,   \label{R0-eq}
\end{equation}
where $g(N)>0$ is a size-dependent function defined similarly to \rref\cite{gonzalez2018exotic} which reads
\begin{equation}
    g(N) = \frac{1}{N(N+1)} \sum_{k,q} \frac{\sin^2(q)\left(2\cos(k/2)\right)^2}{|\omega(k,q)|^4}\,.
\end{equation}
This shows that no fractional decay occurs in our system for any value of $N$.
Importantly, this conclusion crucially depends on the presence of a flat band of edge modes which the qubit is coupled to.
Indeed, 
setting $\Omega=0$ in \eq\eqref{R0-eq} (zero interaction with the flat band) would make $R_0$ finite.
Thus, remarkably, the emergent cavity mode due to the flat band completely washes out the fractional decay.
To further corroborate this conclusion, in \ref{fig:supp_NM}(a), we show the decay of a qubit coupled to a $b$-resonator of the lattice edge [\cf \ref{fig:fig1}(a) in the main text].
Fractional decay now takes place with $P_e\propto g/J$, which is due to the fact that the emergent cavity mode has zero amplitude on the $b$-sublattice [see \ref{fig:supp_NM}].

\subsubsection{Long-time non-exponential decay}

To investigate the long-time decay in the infinite lattice, we need to study the contribution to $c_e(t)$ [\cf\eq\eqref{cet-eq}] coming from the branch cut on the imaginary $i[-\infty, 0]$ interval of the complex plane.
In the long-time limit ($t\to+\infty$), this branch-cut contribution $c_{\rm BC}(t)$ can be expressed as
\begin{equation}
    c_{\rm BC}(t) \simeq -\frac{1}{2\pi} \int_0^\infty \mathrm{d} y \, e^{-yt} \left[G_e(-iy+0^+) - G_e(-iy-0^+)\right]\,,
\end{equation}
where the Green's function, expanded around the origin of the complex plane, can be written as
\begin{equation}
    G_e(-iy \pm 0^+) \simeq -\left\{ iy + \Delta + i \left[ \frac{g^2}{2\pi\sqrt{3}J} \left( \frac{y}{J} \ln \left( \frac{y}{J} \right)^2 \mp i\pi \frac{y}{J} \right) - \frac{\Omega^2}{y} \right] \right\}^{-1} \,.
\end{equation}
Expanding the integrand around $y = 0$, we are able to derive the asymptotic scaling of such contribution, which reads
\begin{equation}
    c_{\rm BC}(t) \simeq \left(\frac{g}{J}\right)^2 \frac{\sqrt{3}}{\pi} \frac{1}{(\Omega t)^4}\,.
\end{equation}
Thus, we predict a power-law decay scaling as $\sim t^{-4}$, which contrasts with the logarithmic scaling observed in the bulk \cite{gonzalez2018exotic}.
Again, this difference is due to a non-zero $\Omega$, which sets a timescale for this non-Markovian decay.
One can easily check that the same scaling law holds for an anisotropic lattice such that $\beta<2$.
Based on above, we thus conclude that at short times, one can neglect the branch-cut contribution and safely make the \textit{single-pole} approximation $\left[z-\Delta-\Sigma_e(\Delta+i0^+)\right]^{-1}$ around $\Delta{=}0$, which is completely equivalent to the Born-Markov approximation \cite{lambropoulos_fundamental_2000} and, as such, yields the same dynamics as the master equation we derived earlier.
At long times such that $t\gg1/\Omega$ (corresponding to several vacuum Rabi oscillations), the qubit population will feature an additional power-law decay going as $P_e(t)=|c_e(t)|^2\!\sim \!(\Omega t)^{-8}$.
This entails that several Rabi oscillation cycles can take place before the branch-cut contribution gives measurable effects.
Indeed, in realistic experimental setups, the required timescale needed to observe this effect could be much longer than the typical timescale of other non-idealities such as decoherence.
Accordingly, the salient features of the decay dynamics are effectively captured by the dissipative Jaynes-Cummings model derived above.
\section{Matrix elements of the decay rate matrix}
\label{sec:gamma}

In this section, we focus on the computation of the matrix elements of matrix $\gamma$.
For the sake of simplicity, we restrict ourselves to the case $\beta\leq 2$ i.e., when the FB is not gapped from the rest of the spectrum.
\subsection{Calculation of $\gamma_{ii}(\Delta)$}
\label{subsec:gammaii}

In the thermodynamic limit, one can express the single qubit decay rate or, equivalently, the element $\gamma_{ii}(\Delta)$ in the dissipation matrix arising from the two qubit master equation as
\begin{equation}
    \frac{\gamma_{ii}(\Delta)}{g^2} = \frac{4}{\pi} \int_{0}^\pi \mathrm{d} q \int_{-\pi}^\pi \mathrm{d} k \frac{\cos^2{(k/2)}\sin^2{(q)}}{\beta^2 +4\beta\cos{(k/2)}\cos{q} + 4\cos^2{(k/2)}}\sum_{\nu}\delta(\Delta-\omega_\nu(k,q))\,.
\end{equation}
Now, since 
\begin{equation}
    \sum_{\nu} \delta(\Delta-\omega_\nu(k,q)) = 2 |\Delta|\, \delta(\Delta^2-J^2\left[\omega(k,q)\right]^2)\,,
\end{equation}
one can simplify the $\nu$ dependency and also to conclude that the denominator is proportional to $(\Delta/J)^2$.
Thus
\begin{equation}
     \frac{\gamma_{ii}(\Delta)}{g^2}= \frac{16J^2}{\pi |\Delta|}\int_{0}^\pi \mathrm{d} q \int_{0}^\pi \mathrm{d} k \cos^2{(k/2)}\sin^2{(q)}\,\delta(\Delta^2-J^2\left[\omega(k,q)\right]^2)\,.
\end{equation}
Changing $x=\cos{q}$ and $y = \cos{k/2}$, we get that 
\begin{equation}
    \begin{split}
         \frac{\gamma_{ii}(\Delta)}{g^2} &= \frac{32 J^2}{\pi|\Delta|} \int_{-1}^1 \mathrm{d} x \int_0^1 \mathrm{d} y \frac{y^2}{\sqrt{1-y^2}}\sqrt{1-x^2} \,\delta\left(4\beta J^2\,y \left(x-\left[\frac{\Delta^2/J^2-\beta^2-4y^2}{4\beta y}\right]\right)\right) = \\
        &= \frac{8}{\beta\pi|\Delta|} \int_{-1}^1 \mathrm{d} x \int_0^1 \mathrm{d} y \frac{y}{\sqrt{1-y^2}}\sqrt{1-x^2} \,\delta\left(x-\left[\frac{\Delta^2/J^2-\beta^2-4y^2}{4\beta y}\right]\right) =\\
        &=\frac{2}{\beta^2\pi|\Delta|} \int_{c_0}^{c_1} \mathrm{d} y \sqrt{\frac{16\beta^2y^2 - \left[\Delta^2/J^2-\beta^2-4y^2\right]^2}{1-y^2}}\,,
    \end{split}
\end{equation}
where $c_0, c_1$ depends on the values of $|\Delta|/J$ and $\beta$.
In particular, one finds that
\begin{equation}
    \begin{split}
        0<\frac{|\Delta|}{J}\leq 1&: 
        \begin{cases}
            \beta<|\Delta|/J & c_0 = (|\Delta|/J-\beta)/2, \quad c_1 = (|\Delta|/J+\beta)/2\\
            |\Delta|/J\leq \beta\leq 2-|\Delta|/J & c_0 = (\beta - |\Delta|/J)/2, \quad c_1 = (|\Delta|/J+\beta)/2\\
            \beta>2-|\Delta|/J & c_0 = (\beta-|\Delta|/J)/2, \quad c_1 = 1\\
        \end{cases}\\
      \frac{|\Delta|}{J}> 1&: 
        \begin{cases}
            \beta<2-|\Delta|/J & c_0 = (|\Delta|/J-\beta)/2, \quad c_1 = (|\Delta|/J+\beta)/2\\
            2-|\Delta|/J\leq \beta\leq |\Delta|/J & c_0 = (|\Delta|/J-\beta)/2, \quad c_1 = 1\\
            \beta>|\Delta|/J & c_0 = (\beta-|\Delta|/J)/2, \quad c_1 = 1\\
        \end{cases}
    \end{split}
    \,.
\end{equation}
Since we are interested in the dynamics of qubits quasi-resonant to the FB, we will focus in the following in the case $0<|\Delta|/J\leq 1$.
In this limit, provided that $ |\Delta|/J\leq \beta\leq 2-|\Delta|/J$, we can simplify the integral bounds by making the substitution $y=\frac{\beta+|\Delta| t/J}{2}$ from which
\begin{equation}
    \frac{\gamma_{ii}(\Delta)}{g^2} = \frac{2|\Delta|}{\pi \beta^2 J^2}\int_{-1}^1 \mathrm{d} t \sqrt{\frac{\left[1-\left(J\frac{1-\beta}{\Delta} + t\right)^2\right]\left[(1+\beta+\Delta t/J)^2 - (\Delta/J)^2\right]}{\left(2+\beta+\Delta t/J\right)\left(2-\beta-\Delta t/J\right)}}\,.
\end{equation}
For $\Delta \simeq 0$, we can approximate the integral, yielding
\begin{align}
    \frac{\gamma_{ii}(\Delta)}{g^2} & \simeq \frac{4 |\Delta| }{\beta\pi J^2 \sqrt{ 4 - \beta^2}}   \int_{-1}^{1} \mathrm{d} t \sqrt{ 1 - t^2} = \frac{2 |\Delta|}{\beta J^2 \sqrt{ 4 - \beta^2}}\,.
\end{align}
Setting $\beta=1$ yields the expression of the single qubit decay rate discussed in the main text.
Notice that this expression would yield a divergent decay rate for $\beta \to 0^+$ or $\beta\to2^-$.
These two limits, though, are out of the validity region of such approximation for any non-zero value of $\Delta$.
As a matter of fact, both limits have clear physical interpretation.
For instance, if $\beta = 0$ then the Hamiltonian reduces to a collection of 1D homogeneous chains (stretching along the $\mathbf{e}_2$ direction), parametrized by $n$.
In this case, we expect $\gamma_{ii}$ to reduce to the standard decay rate into a cosine band.
This is easily checked as in this case $\omega(k,q) = 2J\cos{(k/2)}$ and so
\begin{equation}
    \begin{split}
    \frac{\gamma_{ii}(\Delta)}{g^2} &= \frac{1}{\pi} \int_{-\pi}^\pi \mathrm{d} k \left[\delta(\Delta-2J\cos{(k/2))} + \delta(\Delta+2J\cos{(k/2))}\right]\int_{0}^\pi \mathrm{d} q \sin^2{(q)} = \\
    &= \frac{1}{2}\int_{-\pi}^\pi \mathrm{d} k \left[\delta(\Delta-2J\cos{(k/2))} + \delta(\Delta+2J\cos{(k/2))}\right] \equiv \int_{-\pi}^\pi \mathrm{d} k\, \delta(\Delta-2J\cos{(k)})\,,
    \end{split}
\end{equation}
yielding the expected result according to Fermi Golden Rule.
On the other hand, in the $\beta = 2$ limit, the Dirac cone spectral shape is modified becoming a so--called \textit{semi-Dirac} cone \cite{redondo2021quantum}, whose dispersion law is linear in one quasi-momentum component and quadratic in the other.
For this reason, we expect $\gamma_{ii}$ to be smaller in this case than its value for $\beta<2$.
This intuition is verified by numerically computing the integral and it is shown in Fig. 3(e) in the main text.
Interestingly, we see that the decay rate goes to zero when the qubit is tuned on resonance with the Dirac cone i.e., when $\Delta=0$.
This is understood noticing that, within the Markov approximation, the decay rate is proportional to the local density of states of bulk modes at the bare qubit frequency i.e., to the number of bulk modes at frequency $\Delta$ overlapping the cavity to which the qubit is locally coupled.
Since in graphene the bulk density of states goes to zero on resonance with the Dirac points \cite{Girvin_Yang_2019,economou_greens_1979}, Markov approximation predicts no decay into bulk modes at this frequency.
\subsection{Calculation of $\gamma_{i\not=j}(\Delta)$}
\label{subsec:gammaij}

In the same way as before, one would like to compute 
\begin{equation}
    \frac{\gamma_{ij}(\Delta)}{g^2} = \frac{4}{\pi} \int_{0}^\pi \mathrm{d} q \int_{-\pi}^\pi \mathrm{d} k \frac{\cos^2{(k/2)}\sin^2{(q)}}{\beta^2 +4\beta\cos{(k/2)}\cos{q} + 4\cos^2{(k/2)}}\cos{k(m_i-m_j)}\sum_{\nu}\delta(\Delta-\omega_\nu(k,q))\,.
\end{equation}
Doing the same simplifications as before, this integral is written as
\begin{equation}
     \frac{\gamma_{ij}(\Delta)}{g^2}= \frac{16J^2}{\pi |\Delta|}\int_{0}^\pi \mathrm{d} q \int_{0}^\pi \mathrm{d} k \cos^2{(k/2)}\sin^2{(q)}\cos{(k|m_i-m_j|)}\,\delta(\Delta^2-J^2\left[\omega(k,q)\right]^2)\,.
\end{equation}
The previous calculation suggests to use the substitution $x=\cos q$ and $y = \cos{k/2}$.
In order to do so, one recalls that
\begin{equation}
    \cos{n\theta} = \sum_{r=0}^{2r\leq n} (-1)^r \binom{n}{2r} \cos^{n-2r}{(\theta)} \sin^{2r}{(\theta)}
\end{equation}
and using this, it is possible to write
\begin{equation}
    \begin{split}
         \frac{\gamma_{ij}(\Delta))}{g^2} =\frac{2}{\beta^2\pi|\Delta|} &\sum_{r=0}^{r\leq |m_i-m_j|} (-1)^r \binom{2|m_i-m_j|}{2r}\times\\
         &\int_{c_0}^{c_1} \mathrm{d} y \sqrt{16\beta^2y^2 - \left[\Delta^2/J^2-\beta^2-4y^2\right]^2} \left(1-y^2\right)^{r-1/2} y^{2|m_i-m_j|-2r} 
    \end{split}
\end{equation}
with the same prescription about the integral limits as before.
For $\Delta \simeq 0$, the integral can be approximated as
\begin{equation}
    \begin{split}
     \frac{\gamma_{ij}(\Delta)}{g^2} & \simeq \frac{4 |\Delta|}{\beta\pi J^2 \sqrt{4 - \beta^2}}  \sum_{r = 0}^{ r \leq |m_i - m_j|} (-1)^r \binom{2 |m_i - m_j|}{2 r} \left( \frac{\beta}{2}\right)^{2|m_i - m_j| -2 r} \left( 1-\left(\frac{\beta}{2}\right)^2 \right)^{ r}  \int_{-1}^{1} \mathrm{d} t \sqrt{1 - t^2 } = \\
     & = \frac{2 |\Delta|}{\beta J^2 \sqrt{4 - \beta^2}} \cos{\left(k_D |m_i - m_j|\right)} \,.
\end{split}
\end{equation}
Even in this case we see that around the Dirac point, the scaling of the decay rate is linear in $|\Delta|$.
Notice that in general, the matrix elements of the $\gamma$ matrix depends on $\beta$ in a non-monotonic way.
The decay rates reach a minimum in correspondence of $\beta=\sqrt{2}$, in which case dissipation will be minimized.
The same comments regarding the $\beta \to 0^+$ and $\beta \to 2^-$ limits present at the end of \autoref{subsec:gammaii} apply here.
Interestingly, notice that if $k_D|m_i-m_j| = \nu 2\pi$, where $\nu\in\mathbb{N}$, then $\gamma_{ij} = \gamma_{ii}$ i.e., the off-diagonal matrix element of the decay matrix will reduce to the corresponding diagonal one when $\Delta\simeq0$.
Recalling that $k_D = 2\arccos{\left(\beta/2\right)}$ and that $0<\beta<2$, this happens for values of $\beta=2\cos{\left(\nu\pi/|m_i-m_j|\right)}$ for $0<\nu<|m_i-m_j|/2$.
Thus, the minimum value for which a non-trivial solution exists is $|m_i-m_j|=3$, yielding $\beta=1$.
Once this condition is verified for a value $\Delta m$ of the distance between qubits, the decay rate of every pair of qubits placed at a distance which is an integer multiple of $\Delta m$ will share the same property.
\section{Exact solution of the dissipative Jaynes-Cummings model}

The master equation (5) reported in the main text describes a dissipative Jaynes-Cummings model in which --unlike standard cases-- a \textit{lossy} qubit couples to a \textit{lossless} bosonic cavity.
In a different context, the same model was already considered in \rref \cite{lorenzo_quantum_2017}, where a complete analytical solution in the single-excitation sector is discussed.
Taking inspiration from this, here we discuss the general solution of our model, depending on the values of $\Delta$ and $\beta<2$.
Suppose that the system is initially in the pure state $\rho = |e\rangle\langle e|$, corresponding to an initially excited qubit.
At each later time, the probability of finding the qubit in its excited state can be found by solving the master equation.
To do so, one notices that the single-qubit master equation can be rewritten as
\begin{equation}
    \dot{\rho} = -i\left(\mathcal{H}_{\rm NH}\, \rho - \rho\mathcal{H}^\dagger_{\rm NH}\right) + \gamma(\Delta)\sigma \rho \sigma^\dagger\,,
\end{equation}
where the last term is the usual recycling term and the non-Hermitian Hamiltonian reads
\begin{equation}
    \mathcal{H}_{\rm NH} = \left(\Delta-i\frac{\gamma(\Delta)}{2}\right)\sigma^\dagger \sigma + \Omega\left(\sigma \mathcal{C}^\dagger +\hc\right)\,.
\end{equation}
Within the single-excitation sector, spanned by states $|e, {\rm vac}\rangle$ and $|g, \mathcal{C}\rangle$, the matrix elements of the recycling term yields no contribution.
This allows to cast the master equation into a Von-Neumann-like equation, where the dynamics is generated by the non-Hermitian Hamiltonian $\mathcal{H}_{\rm NH}$.
Solving this is equivalent to solving the time-dependent Schr\"odinger equation
\begin{equation}
    i\frac{\mathrm{d}}{\mathrm{d}t}|\Psi(t)\rangle = \mathcal{H}_{\rm NH} |\Psi(t)\rangle\,,
\end{equation}
where $\rho(t) = |\Psi(t)\rangle\langle\Psi(t)|$. Writing the state of the system at a generic time $t$ as $|\Psi(t)\rangle = c_e(t) |e,\rm vac\rangle + c_\mathrm{g}(t) |g, \mathcal{C}\rangle$ and solving for $c_e(t)$, we end up with the second order homogeneous linear differential equation
\begin{equation}
    \frac{\mathrm{d}^2}{\mathrm{d}t^2}c_e(t) + \nu \frac{\mathrm{d}}{\mathrm{d}t}c_e(t) + \Omega^2 c_e(t) = 0\,, 
\end{equation}
where $\nu = i\Delta+\gamma(\Delta)/2$ and the initial value conditions are $c_e(0)=1$ and $\dot{c}_e(0) = -\nu$. Taking $c_e(t) = e^{\lambda t}$ as an ansatz for the solution, the characteristic polynomial associated to 
the differential equation reads $p(\lambda) {=} \lambda^2 {+} \nu\lambda {+}\Omega^2$, whose roots are
\begin{equation}
    \lambda_\pm = -\frac{\nu}{2} \pm \frac{1}{2} \sqrt{\nu^2-4\Omega^2}\equiv-\frac{\nu}{2} \pm \frac{i}{2} \sqrt{4\Omega^2-\nu^2}\,.
\end{equation}
This allows us to write the general solution as
\begin{equation}
    c_e(t) = e^{-\nu t/2} \left[\cos{\left(\frac{\sqrt{4\Omega^2-\nu^2}}{2}t\right)} - \frac{\nu}{\sqrt{4\Omega^2-\nu^2}}\sin{\left(\frac{\sqrt{4\Omega^2-\nu^2}}{2}t\right)}\right]\,.
\end{equation}
We now study two particular limits of this expression. In the limit $\Delta\lesssim\Omega$, recalling $\Omega\propto g$ while $\gamma(\Delta)\propto g^2$, we can neglect $\gamma$ with respect to $\Delta$, ending up with 
\begin{equation}
    c_e(t) \simeq e^{-i\Delta t/2} \left[\cos{\left(\frac{\Omega_R}{2}t\right)} - \frac{i\Delta}{\Omega_R}\sin{\left(\frac{\Omega_R}{2}t\right)}\right]\,,
\end{equation}
where $\Omega_R=\sqrt{\Delta^2+4\Omega^2}$, meaning that the qubit population goes as
\begin{equation}
    p_e(t) = |c_e(t)|^2 = 1 - \frac{4\Omega^2}{\Omega_R^2}\sin^2{\left(\frac{\Omega_R}{2}t\right)}\,,
\end{equation}
as expected for a standard Jaynes-Cummings model with no dissipation.
On the other hand, if $\Delta \gg \Omega$ we can neglect $\Omega$ with respect to $\gamma$ meaning that $c_e(t)$ can be approximated as
\begin{equation}
    c_e(t) \simeq e^{-\nu t} = e^{-i\Delta t} e^{-\gamma(\Delta) t/2}\,,
\end{equation}
meaning that the qubit undergoes spontaneous emission at a rate $\gamma(\Delta)$.
For intermediate values of $\Delta$, the qubit population interpolates between these two behaviors, going from Rabi oscillations to exponential decay as $\Delta$ increases.
\section{Two-qubit dynamics in the single-excitation sector}

In this section, we address the dynamics of two qubits coupled a zigzag edge of photonic graphene.
For the sake of clarity, we first consider the resonant case $\Delta=0$ where dissipation into the bulk modes is negligible, and only afterwards address the case $\Delta\not=0$ yielding non-negligible dissipation.
For $\Delta=0$, the effective model would predict a purely-coherent dynamics generated by the Hamiltonian
\begin{equation}
    \label{eq:H-2qubit-single}
    \mathcal{H} = 
    \begin{cases}
        \Omega \left[\left(\sigma_1^\dag + \sigma_2^\dag\right)\mathcal{C}+\hc\right]\,, & \text{if } \mathbf{R}_1=\mathbf{R}_2\,,\\
        g \sum_{i,j=1}^2 \left(\sigma_i^\dagger M_{ij} \mathcal{C}_{j} + \hc\right)\,, & \text{otherwise}\,,
    \end{cases}
\end{equation}
depending on whether the qubits are coupled to the same resonators ($\mathbf{R}_1=\mathbf{R}_2$), in which the Tavis-Cummings model is recovered, or to different resonators, where one needs to introduce two bosonic modes [see \autoref{subsec:many-qubit}].
Here, $M_{ij}$ are the matrix elements of the $2\times2$ matrix $\mathbf{M}$ defined as $\mathbf{M}\mathbf{M}^\dag = \mathbf{P}$, being $\mathbf{P}$ the FB projector computed at the qubits' positions.
While the exact values of coefficients $M_{ij}$ are not relevant for the purpose of the present section, it is  important to note that they show up a power-law dependence on the inter-emitter distances $|m_i-m_j|$.
This property is inherited from the power-law scaling of $P_{ij}$ discussed in \autoref{app:Gij}.
We consider initial states of the form
\begin{equation}
    \label{eq:general-initstate}
    |\Psi(0)\rangle = \cos{\left(\frac{\theta}{2}\right)}|e_1g_2,\rm vac\rangle 
    + e^{i\varphi}\sin{\left(\frac{\theta}{2}\right)}|g_1e_2,\rm vac\rangle\,,
\end{equation}
where $0\leq\theta<\pi$ and $0\leq \varphi<2\pi$.
The evolved state at a generic time $t$ will remain within the single-excitation sector.
If the two qubits are coupled to the same resonator (Tavis-Cummings model), starting from the state in \eqref{eq:general-initstate} the system will evolve into $|\Psi(t)\rangle = c_{e1}(t) |e_1g_2,\rm vac\rangle + c_{e2}(t) |e_1g_2,\rm vac\rangle + d(t) |g_1g_2,\mathcal{C}\rangle$ with
\begin{align}
    c_{e1}(t) = \frac{c_-+c_+\cos{\left(\Omega\sqrt{2}t\right)}}{2}\,, && c_{e2}(t) = \frac{-c_-+c_+\cos{\left(\Omega\sqrt{2}t\right)}}{2}\,, && d(t) = -\frac{i}{\sqrt{2}}c_+ \sin{\left(\Omega\sqrt{2}t\right)}\,,
\end{align}
being $c_{\pm} = \cos{\left(\frac{\theta}{2}\right)} \pm e^{i\varphi}\sin{\left(\frac{\theta}{2}\right)}$, evolution which is derived from the first line of \eq \eqref{eq:H-2qubit-single}.
From this general solution, we can see that if $\theta=0$ (first qubit initially excited), an excitation is periodically exchanged between the two qubits at frequency $\Omega\sqrt{2}$ (collective enhancement).
The antisymmetric combination corresponding to $\theta = -\pi/2$ and $\phi=0$ is instead a dark state, as expected in the single-excitation sector of the Tavis-Cummings model \cite{fink_dressed_2009}.
More generally, if two qubits are coupled to the different resonators, their dynamics is subject to the Hamiltonian given in the second line of \eq\eqref{eq:H-2qubit-single}.
Hence, the form of the general state in the single-excitation sector reads
\begin{equation}
    |\Psi(t)\rangle = c_{e1}(t) |e_1 g_2,{\rm vac}\rangle + c_{e2}(t) |g_1 e_2,{\rm vac}\rangle + d_{1}(t) |g_1 g_2,{\cal C}_1\rangle + d_{2}(t) |g_1 g_2,{\cal C}_2\rangle\,.
\end{equation}
\begin{figure}[t]
    \centering
    \includegraphics[width=0.8\linewidth]{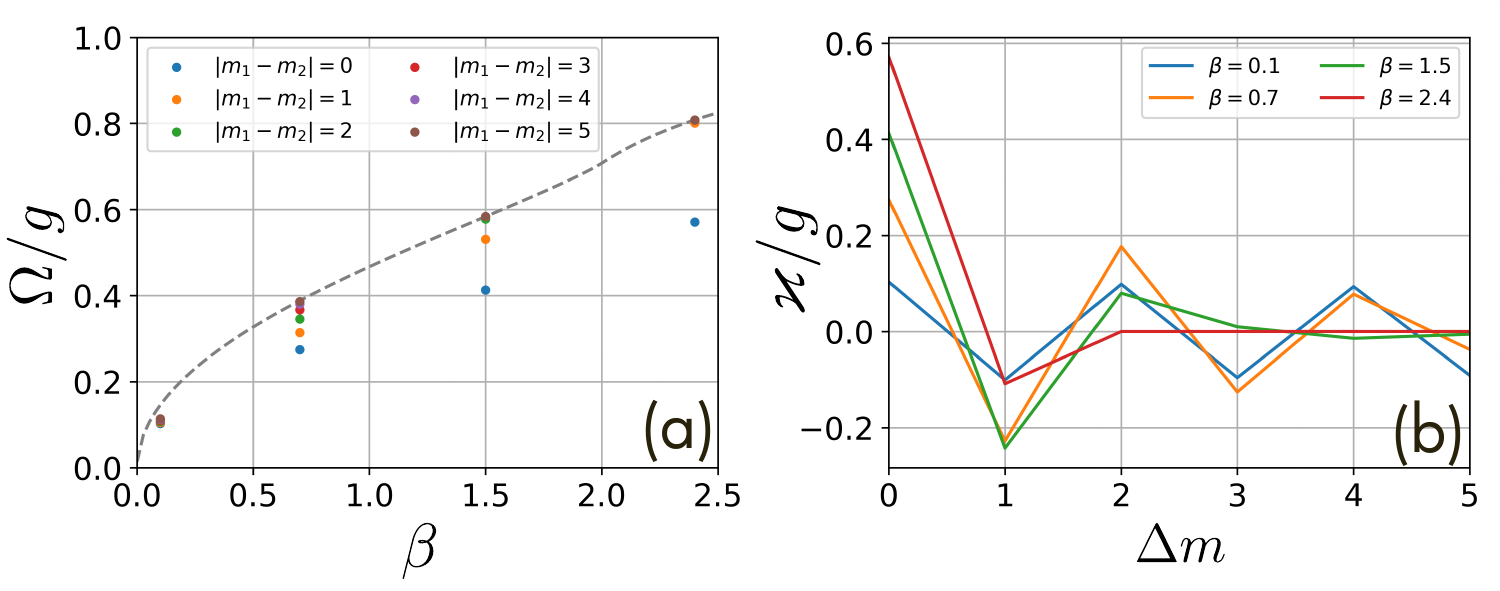}
    \caption{Characteristic frequencies in the resonant state transfer.
(a) Diagonal frequency $\Omega/g=M_{11}$ as a function of the anisotropy parameter $\beta$ for different values of the qubits' spacing $\Delta m=|m_1-m_2|$ (from $0$ to $5$).
The grey dashed line represent the single-qubit Rabi frequency as a function of $\beta$.
(b) Off-diagonal frequency $\varkappa/g=M_{12}$ as a function of the spacing between the qubits for different values of $\beta=0.1,0.7,1.5,2.4$ (same as the points in panel a).}
    \label{fig:omegavarkappa}
\end{figure}
Plugging this into the time-dependent Schr\"odinger equation thus yields the differential system
\begin{equation}\label{diff-s}
    \begin{cases}
        i\frac{\mathrm{d}}{\mathrm{d}t}c_{e1}(t) = g M_{11} d_1 (t) + g M_{12} d_2 (t)\\ 
        i\frac{\mathrm{d}}{\mathrm{d}t}c_{e2}(t) = g M_{21} d_1 (t) + g M_{22} d_2 (t)\\
        i\frac{\mathrm{d}}{\mathrm{d}t}d_1(t) = g M_{11} c_{e1} (t) + g M_{12} c_{e2} (t)\\
  
      i\frac{\mathrm{d}}{\mathrm{d}t}d_2(t) = g M_{21} c_{e1} (t) + g M_{22} c_{e2} (t)\\
    \end{cases}
    \,
\end{equation}
subject to the initial conditions $c_{e1}(0) {=} \cos{\theta/2}$ and $c_{e2}(0) {=} e^{i\varphi} \sin{\theta/2}$ and $d_1(0) {=} d_2(0) {=} 0$. Upon the replacement $c_{ei} = A_i e^{-i\omega t}$ and $d_{i} = \Tilde{A}_i e^{-i\omega t}$, the differential system can be reduced to the eigenvalue problem defined by
\begin{equation}
    \mathcal{M} \vec{A} = \frac{\omega}{g} \vec{A}\,,
\end{equation}
where $\vec{A} = \left(A_1\;\;
A_2\;\;\Tilde{A}_1\;\;\Tilde{A}_2\right)^T$ while the dynamical matrix $\mathcal{M}$ is given by
\begin{equation}
    \mathcal{M} = 
    \begin{pmatrix}
        \mathbf{0}_{2} & M\\
        M^\dagger & \mathbf{0}_{2} 
    \end{pmatrix}
    \,,
\end{equation}
where $\mathbf{0}_2$ is the $2\times2$ zero matrix.
The system's symmetries impose the constraints $M_{11} {=} M_{22}$ and $M_{12} {=} M_{21}$.
This implies that matrix $\mathbf{M}$ is real and Hermitian, hence the dynamical matrix $\mathcal{M}$ has eigenvalues
\begin{align}
        \omega_{1,\pm} = -\Omega \pm \varkappa\,, && \omega_{2,\pm} = \Omega \pm \varkappa\,,
\end{align}
where $\Omega {=} g M_{11}$ is the resonant Rabi frequency while $\varkappa{=}g M_{12}$ originates from the overlap between the two emergent cavity modes seeded by the qubits.
The corresponding eigenstates reads
\begin{align}
    \vec{v}_{1,\pm} = \frac{1}{2} \left(\pm1\;\;-1\;\;\mp1\;\;1\right)^T\,, &&
    \vec{v}_{2,\pm} = \frac{1}{2} \left(\pm1\;\;1\;\;\pm1\;\;1\right)^T\,.
\end{align}
In this way, we can express the solution of the differential system \eqref{diff-s} as a linear combination of oscillatory terms.
Upon taking the squared modulus of each amplitude, the population of the two qubits and cavity modes are worked out as
\begin{equation}
    \begin{cases}
        P_{e1}(t) = \frac{1}{4}\left[1+\cos{\left(2\Omega t\right)}\cos{\left(2\varkappa t\right)} + \cos{\left(\theta\right)}\left(\cos{\left(2\Omega t\right)} + \cos{\left(2\varkappa t\right)}\right) - \cos{\left(\varphi\right)}\sin{\left(\theta\right)}\sin{\left(2\Omega t\right)}\sin{\left(2\varkappa t\right)}\right]\\
        P_{e2}(t) = \frac{1}{4}\left[1+\cos{\left(2\Omega t\right)}\cos{\left(2\varkappa t\right)} - \cos{\left(\theta\right)}\left(\cos{\left(2\Omega t\right)} + \cos{\left(2\varkappa t\right)}\right) - \cos{\left(\varphi\right)}\sin{\left(\theta\right)}\sin{\left(2\Omega t\right)}\sin{\left(2\varkappa t\right)}\right]\\
        P_{d1}(t) = \frac{1}{4}\left[1-\cos{\left(2\Omega t\right)}\cos{\left(2\varkappa t\right)}  -\cos{\left(\theta\right)}\left(\cos{\left(2\Omega t\right)} - \cos{\left(2\varkappa t\right)}\right) + \cos{\left(\varphi\right)}\sin{\left(\theta\right)}\sin{\left(2\Omega t\right)}\sin{\left(2\varkappa t\right)}\right]\\
        P_{d2}(t) = 
\frac{1}{4}\left[1-\cos{\left(2\Omega t\right)}\cos{\left(2\varkappa t\right)} + \cos{\left(\theta\right)}\left(\cos{\left(2\Omega t\right)} - \cos{\left(2\varkappa t\right)}\right) +\cos{\left(\varphi\right)}\sin{\left(\theta\right)}\sin{\left(2\Omega t\right)}\sin{\left(2\varkappa t\right)}\right]
    \end{cases}
    \,.
\end{equation}
%We now discuss a few limits. 
For $\theta=0$, corresponding to the initial state (relevant for $1{\rightarrow} 2$ state transfer) $|\Psi(0)\rangle=|e_1g_2,{\rm vac}\rangle$, the qubit populations read
\begin{align}
    P_{e1}(t) = \cos^2{(\Omega t)}\,\cos^2{(\varkappa t)}\,, && P_{e2}(t) = \sin^2{(\Omega t)}\,\sin^2{(\varkappa t)}\,.
\end{align}
The presence of two characteristic frequencies ($\Omega$ and $\varkappa$) show the origin of the beatings in Fig. 2(d)-(e) of the main text.
Perfect state transfer at a specific time $t=\tau$ occurs when
$P_{e1}(\tau) = 0$ and $P_{e2}(\tau) = 1$.
This enforces $\sin^2{\left(\Omega\tau\right)} = 1$ and $\sin^2{\left(\varkappa\tau\right)} = 1$. The former and latter condition are respectively satisfied at times $\tau_0 = \frac{\pi+2p\pi}{2\Omega}$ and $\tau_1 = \frac{\pi+2q\pi}{2\varkappa}$ with $p,q\in \mathbb{Z}$.
Accordingly, perfect state transfer is reached if and only if $\tau_0 = \tau_1$, which is equivalent to the condition
\begin{equation}
    p = \frac{\Omega}{\varkappa} q + \frac{1}{2} \left(\frac{\Omega}{\varkappa}-1\right)\,.
\end{equation}
This equation has an integer solution only if the ratio between $\Omega$ and $\varkappa$ is commensurate, i.e., for $\Omega/\varkappa\in2\mathbb{Z}+1$ (odd integers).
Thus, although dissipation is negligible, perfect state transfer (corresponding to 100\% fidelity) may still not be attainable.
Nonetheless, in typical cases one can achieve very high fidelity, as in the instance of Fig. 2(d-e) of the main text where fidelity reaches $\simeq 94\%$ for $m_2-m_1=2$.
In \ref{fig:omegavarkappa}, we plot the two characteristic frequencies $\Omega$ and $\varkappa$ as functions of the relevant parameters, in this case the qubits' spacing $\Delta m$ and the anisotropy parameter $\beta$.
We see that the farther apart the qubits are, the more $\Omega$ tends to the single-qubit resonant Rabi frequency, signaling an effective decoupling of the dynamics of the two qubits one another.
The decaying profile of $\varkappa$ can be tied to the decaying profile of the effective cavity modes.
Similar to them, $\varkappa$ has a hard cutoff for $\Delta m\geq 2$ if $\beta>2$, while decaying more slowly the smaller the value of $\beta$.

For $\Delta\not=0$, losses into the bulk modes are no longer negligible making state transfer intrinsically dissipative (i.e. part of the initial qubit excitation unavoidably leaks into the bulk [see Fig. 2(d)-(e) of the main text]).
Interestingly, around $\Delta\simeq0$, the decay rate matrix $\gamma_{ij}$ is proportional to $\cos{\left(k_D|m_i-m_j|\right)}$ [\cf \autoref{app:Gij} and \autoref{subsec:gammaij}].
Thus, according to the discussion in \autoref{subsec:gammaij}, for any value of $\Delta m=|m_i-m_j|\geq 3$ one can find at least one value of $\beta$ for which the decay matrix will be proportional to a constant matrix, namely $\gamma_{ij}=\gamma_0$.
In such cases the dissipator in the master equation \eqref{eq:ME2} can be recast so as to feature only one \textit{collective} jump operator $J=\sigma_1+\sigma_2$
\begin{align}
    \mathcal{L}[\rho] = \gamma_0 \left(J\rho J^\dag - \frac{1}{2}\{J^\dag J, \rho\} \right)\,, &&\text{with}\quad\quad \gamma_0 = \frac{g^2|\Delta|}{J^2 \beta\sqrt{4-\beta^2}}\,.
\end{align}
This is reminiscent of the dissipative Dicke model \cite{gelhausen2018dissipative}.
Once we ascertained that this is the form of the dissipator, it is easy to show that the symmetric superposition $|e_1g_2\rangle+|g_1e_2\rangle$ will dissipate at rate $2\gamma_0$ while the antisymmetric state $|e_1g_2\rangle-|g_1e_2\rangle$ will be a subradiant state fully protected from dissipation.
This result can be straightforwardly extended to more than two qubits.
While perfect loss-free state transfer is indeed impossible in the $\Delta\not=0$ case, engineering the condition $\gamma_{ij} = \gamma_0$ (Dicke-like dynamics) provides an optimal upper bound for the transfer efficiency, as the permanent preservation of the subradiant component guarantees the fidelity remains strictly higher than the case of independent individual decay.

\section{Flat band projection operator: relevant matrix elements }\label{app:Gij}

The projector onto the FB subspace is defined as the operator $\mathbf{P} = \int \mathrm{d} k |\mathcal{E}_k\rangle\langle\mathcal{E}_k|$, where $k$ runs over the region in \eq\eqref{eq:support} and with $|\mathcal{E}_k\rangle = \mathcal{E}_k^\dagger|{\rm vac}\rangle$ is the single-photon edge states corresponding to edge modes.
Setting and $k_D {=} 2\arccos{(\beta/2)}$, the relevant matrix elements of the flat band projector between the $a$-resonators on the zigzag edge at positions $m_i$ and $m_j$(i.e., those relevant to our study) are
\begin{equation}\label{eq:int4}
    P_{ij}(\beta) = \langle a_{0\,m_i} | \mathbf{P} | a_{0\,m_j} \rangle = \frac{1}{\beta^2\pi}\int_{k_D}^{\pi} \mathrm{d} k \left(\beta^2-2-2\cos{k}\right) \, \cos{\left[k(m_i-m_j)\right]}\,.
\end{equation}
Setting $m{=}|m_i-m_j|$, the integral can be calculated exactly and for $\beta<2$ we end up with
\begin{equation}
    \label{eq:non-compactFB}
    P_{ij}(\beta<2) = \frac{1}{\beta^2\pi}\times
    \begin{cases}
        2\sin{(k_D)}+(\beta^2-2)\left(\pi-k_D\right) & {\rm for}\,\,m=0\\
        (1-\beta^2/2)\sin{(k_D)}-2\arcsin{(\beta/2)} &{\rm for}\,\, m=1\\
        (2-\beta^2) s_\beta(m) + s_\beta(m+1) + s_\beta(m-1) &{\rm for}\,\,|m|\ge2
    \end{cases}\,,
\end{equation}
where $s_\beta(x) = \sin{(k_Dx)}/(k_Dx)$ is a cardinal sine function with a rescaled argument, depending on $k_D$, hence on $\beta$.
\noindent
For $\beta\geq 2$, the lower bound of integral \eqref{eq:int4} vanishes and we obtain
\begin{equation}
    \label{eq:compactFB}
    P_{ij}(\beta\geq2) = 
    \begin{cases}
        1-2/\beta^2 & {\rm for}\,\,m=0\\
        -1/\beta^2 & {\rm for}\,\,m=1\\ 
        0 &{\rm for}\,\,|m|\ge 2
    \end{cases}
    \,.
\end{equation}
Thus, interestingly, $P_{ij}$ scales as the inverse of the distance $m$ between emitters for $\beta<2$, while it becomes  strictly compact for $\beta\geq 2$.
Notice that $P_{ii}(\beta)$ corresponds to the integral in \eq \eqref{eq:commutator}, for which reason the normalization constant in the effective cavity modes introduced before can be in general expressed as $\mathcal{A} = 1/P_{ii}(\beta)$.
Furthermore, as it appears evident from the form of the cavity mode, we have that $c(0,m_i-m_j) = P_{ij}(\beta)$.
Thus, $P_{ij}(\beta)$ also represents the shape of the effective cavity mode computed on the zigzag edge.
In the following, we leverage this knowledge in order to discuss the scaling at long distances of the cavity modes, both on the edge and toward the bulk.
\section{Shape of the effective cavity modes}
\label{sec:scalingC}

As discussed in the main text, each qubit coupled to the edge of graphene seeds a corresponding cavity mode.
Despite the complexity of its analytic form, the mode's shape at long distances can be analyzed to highlight its scaling properties.
We start by looking at the shape of the cavity mode along the zigzag edge, namely the $m$-dependent function $c(0,m)$.
%for $m\gg1$. 
Recalling that this quantity is equivalent to the FB projector, for $m\geq2$ it can be expressed as
\begin{equation}
    c(0,m) = \frac{1}{\pi\beta^2}\left[ (2-\beta^2)\frac{\sin{k_Dm}}{m} + \frac{\sin{k_D(m+1)}}{m+1} +\frac{\sin{k_D (|m|-1)}}{m-1}\right]\,,
\end{equation}
where $k_D=2\arccos{(\beta/2)}$.
For $m \gg 1$, one can make the expand
\begin{equation}
    \frac{1}{|m|\pm1} = \frac{1}{|m|}\left( \frac{1}{1\pm |m|^{-1}}\right) = \frac{1}{|m|} \left[1 \mp \frac{1}{|m|}\right] + o\left(m^{-2}\right)\,.
\end{equation}
Using that $\sin{k_D(m\pm1)} = \sin{(k_D m)}\cos{k_D} \pm \cos{(k_D m)}\sin{k_D}$, we thus get that the cavity mode along the edge scales as $m^{-2}$ according to
\begin{equation}
    c(0,m\gg1) = -\frac{\sqrt{4-\beta^2}}{\pi\beta}\frac{\cos{(k_D m)}}{m^2} + o\left( m^{-2}\right)\,.
\end{equation}
Accordingly, the projector shows up the same scaling, $P_{ij} \propto |m_i-m_j|^{-2}$.
The computation of the scaling along the edge-bulk direction is somewhat more involved.
The amplitude of the effective cavity [\cf \eq (6)] on resonator $a_{nm}$ is given by
\begin{equation}
    \begin{split}
    c(n,m) &= \frac{1}{\sqrt{2\pi}}\int_{k_D{<}|k|{<}\pi}\!\mathrm{d} k \,{\cal N}_k\,\varepsilon_k(n,m) \\
    &= \frac{(-1)^{n}}{\pi\beta^2}\int_{k_D}^\pi\!\mathrm{d} k\, (\beta^2-2-2\cos{k}) \,\left[\frac{2}{\beta}\cos{\left(\frac{k}{2}\right)}\right]^n\, \cos{\left(k(n/2+m)\right)}
    \end{split}
\end{equation}
where $\mathcal{N}_k = \sqrt{-(2/\beta^2-1+(2/\beta^2)\cos{k})}$ and $\varepsilon_k(n,m)$ is the amplitude of edge mode $\mathcal{E}_k$ on the same resonator.
The cavity mode is represented here as a superposition of edge modes.
These modes are characterized by a localization length $\lambda_k$ that varies continuously from zero at $k = \pi$ to a divergence at the critical point $k = k_D$.
Clearly, the dominant contribution to $c(n\gg1,m)$ comes from the edge modes with the largest localization length corresponding to $k=k_D$.
Accordingly, we can expand the integrand around $k_D$ to leading order.
Thus, using that
\begin{equation}
    \begin{split}
        \beta^2-2-2\cos{k} &= \beta\sqrt{4-\beta^2} \epsilon + o(\epsilon)\,,\\
        \frac{2}{\beta}\cos{\left(\frac{k}{2}\right)} &= 1-\frac{\sqrt{4-\beta^2}}{2\beta}\epsilon + o(\epsilon) \equiv 1-\alpha\epsilon+o(\epsilon) \,\Rightarrow\, \left[\frac{2}{\beta}\cos{\left(\frac{k}{2}\right)}\right]^n \simeq e^{-n\alpha\epsilon}\,,
    \end{split}
\end{equation}
with $\epsilon=k-k_D$, we end up with
\begin{equation}
    \begin{split}
        c(n\gg 1,m) &\simeq \frac{(-1)^n}{\pi\beta}\sqrt{4-\beta^2} \cos{\left(k_D(n/2+m)\right)}\int_0^\infty \mathrm{d} \epsilon\, \epsilon e^{-n\alpha\epsilon} \\
        &= (-1)^n\frac{ 4\beta}{\pi\sqrt{4-\beta^2}} \frac{\cos{\left(k_D(n/2+m)\right)}}{n^2}\,,
    \end{split}
\end{equation}
where we extended the range of integration to the interval $[0, \infty)$ (due to the decreasing nature of 
the exponential) and substituted the value of $\alpha$. We thus see that the leading term for $n\gg 1$ scales as $n^{-2}$.
We thus conclude that the emergent mode amplitude scales as a power law with an exponent of $-2$, both along the edge and toward the lattice bulk.

\begin{figure}[t]
    \centering
    \includegraphics[width=0.8\linewidth]{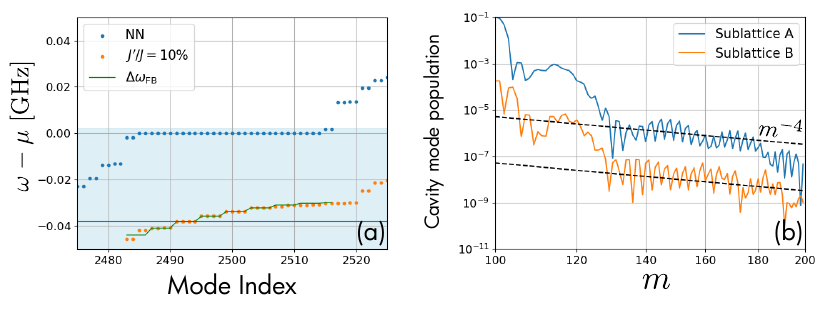}
    \caption{Zigzag graphene in presence of next-nearest-neighbor (NNN) interactions, $J'$. (a) Spectrum for a $50\times50$ unit cells system, with nearest-neighbor hopping $J$ equal to $150\si{\mega\hertz}$. Blue (orange) points correspond to the spectrum of systems with $J'/J = 0\%$ ($J'/J = 10\%$). The blue shaded area around the center of the shifted FB has a width corresponding to the qubit-CCA coupling rate used in panel (b) i.e., $g/2\pi = 40\si{\mega\hertz}$. The green line correponds to the perturbative correction computed in \eq \eqref{eq:correction_NNN}. (b) Effective cavity mode. Simulations are made for a system comprising $200\times200$ unit cells, with $J/2\pi = 150 \si{\mega\hertz}$, $J'/J = 10\%$ and $g/2\pi = 40\si{\mega\hertz}$. The qubit is coupled to the center of the zigzag edge (hence, $m=100$). The blue (orange) line shows the population of edge resonators belonging to the A (B) sublattice. At long distances, both exhibit the same power-law scaling, consistent with the expected $m^{-4}$ scaling.}
    \label{fig:NNN}
\end{figure}

We now investigate whether this scaling survives in the presence of next-nearest-neighbor (NNN) hoppings, which inevitably break chiral symmetry. First, note that the inverse-square power-law scaling originates from the existence of a sharp momentum cutoff at $k=k_D$ in the edge states' wavefunction. If this cutoff also exists in the presence of NNN hoppings, we can claim that the long-distance scaling remains intact. To demonstrate this, we examine where the two bulk bands close their gap under fully periodic boundary conditions (i.e., the position of the Dirac cones), as this corresponds to the position of the momentum cutoff. We note that NNN hoppings do not break translation symmetry and, importantly, connect only sites belonging to the same sublattice. Thus, under periodic boundary conditions along both directions, the Bloch Hamiltonian can be written as
\begin{equation}
    H(k,q) =
    \begin{pmatrix}
        V(k,q) && H_{ \rm AB}(k,q)\\
        H^*_{\rm AB}(k,q) && V(k,q)
    \end{pmatrix}
\end{equation}
where $H_{ \rm AB}(k,q)$ is the off-diagonal term describing nearest-neighbor hoppings, and $V(k,q)$ describes the action of NNN hoppings, which act identically on both sublattices due to sublattice symmetry. Clearly, the eigenvalues of this matrix are $E_{\pm}(k,q) = V(k,q) \pm  \vert{}H_{\rm AB}(k,q)\vert{}$. Thus, a gap closing occurs where $\vert{}H_{\rm AB}(k,q)\vert{}=0$, which is the exact condition derived for the system \textit{without} NNN hoppings. Consequently, the same sharp momentum cutoff is present, and the same scaling is observed.Alternatively, under the assumption of a \textit{weak} NNN coupling ($J'\ll J$), we can use first-order perturbation theory to compute the corrections to the energy and states of the edge modes. First, we define the Hamiltonian describing NNN hoppings as $H_{\rm NNN} = V_A + V_B$, where $V_A$ ($V_B$) describes hoppings between sites in the $a$- ($b$-) sublattice. Since $H_{\rm NNN}$ does not break translation invariance, this perturbation only mixes states with the same $k$ (that is, $k$ remains a good quantum number). We can explicitly write
\begin{align}
    V_A = J'\sum_{n=0}^\infty \left[2\cos(k) a_{kn}^\dagger a_{kn} + 2\cos\left(\frac{k}{2}\right)\left(a_{kn+1}^\dagger a_{kn} + \hc\right)\right]\,,\\
    V_B = J'\sum_{n=0}^\infty \left[2\cos(k) b_{kn}^\dagger b_{kn} + 2\cos\left(\frac{k}{2}\right)\left(b_{kn+1}^\dagger b_{kn} + \hc\right)\right]\,,
\end{align}
and compute the energy correction of the edge modes as ($\mu=0$)
\begin{equation}
    \label{eq:correction_NNN}
    \Delta \omega_{\rm FB}(k) = 2J'\cos(k) - \frac{4J'}{\beta}(1 + \cos k)\,.
\end{equation}
The correction to the states instead reads
\begin{equation}
    |\mathcal{E}_k^{(1)} \rangle \simeq |\mathcal{E}_k\rangle +
        \frac{J'}{J}\mathcal{N}_k
        \sum_{n=0}^\infty (-1)^{n+1} e^{-(n+1)/\lambda_k}|b_{kn}\rangle\,.
\end{equation}
Thus, in the presence of weak NNN hoppings, the edge states acquire a non-zero amplitude on the $b$-sublattice while keeping their amplitude on the $a$-sublattice unchanged. Importantly, the amplitude on the $b$-sublattice decays with the same penetration length as the unperturbed edge modes. As a result, the scaling of the cavity mode remains unchanged, although it now features a non-zero amplitude on the $b$-sublattice as well.
We verified both \eq \eqref{eq:correction_NNN} and the scaling of the corrected cavity mode via numerical simulations, reported in \ref{fig:NNN}. We see that, even for values of $J'\simeq 0.1 J$, the agreement between the correction to the energy and the actual spectrum is remarkable [see \ref{fig:NNN}(a)]. Additionally, we see from \ref{fig:NNN}(b) that the population of the cavities on the edge in both sublattices possesses the same $m^{-4}$ scaling, consistent with the perturbative treatment above.

\section{Qubit-qubit interaction potential in the dispersive regime}
\label{sec:Kij}

The dispersive regime is reached when the detuning of the qubits' frequency and the FB is much bigger than the interaction strength i.e., $|\Delta|\gg \Omega$.
Although achievable, for $\mu=0$ this regime would feature enhanced dissipation into the bulk, spoiling completely the coherent dynamics mediated by the cavity photons.
To circumvent this issue, we set $\mu\not=0$, thus opening a gap in the middle of the spectrum.
This shifts the FB frequency to $\mu$ but, at the same time, cancels dissipations in the energy interval $[-\mu, \mu]$ (within the Markov approximation) corresponding to the bandgap.
In this way, upon tuning the qubits' frequency into the bandgap, we reach the dispersive regime while avoiding any leak of the qubits' population in the bulk.
The ensuing dynamics is thus completely coherent.

Thus, assuming both $|\Delta|<\mu$ and $|\Delta-\mu|\gg \Omega$ (dispersive condition) and considering only the case in which the qubits are coupled to the edge, we can expand the interaction Hamiltonian of the full system up the second order in perturbation theory \cite{bravyi_schriefferwolff_2011}, yielding the effective Hamiltonian (valid provided that the bath is initially in its vacuum state) $\mathcal{H}_{\rm eff} = \sum_{ij} \mathcal{K}_{ij}\sigma_i^\dagger \sigma_j$, where the interaction potential $\mathcal{K}_{ij}$ reads
\begin{equation}
    \frac{\mathcal{K}_{ij}}{g^2} = \frac{1}{\Delta-\mu} c(0,|m_i-m_j|) + G_{\rm bulk}(\Delta)\,,
\end{equation}
where $c(0,m)$ is the cavity mode wavefunction (equivalent to the matrix element of the FB projector) 
[see \autoref{app:Gij}] and $G_{\rm bulk}(\Delta)$ is the matrix element of the bulk modes' Green's operator, computed at the qubit frequency $\Delta$, computed between the two resonators to which qubit $i$ and $j$ are coupled to.
Although being a 2D system, for which the general theory developed in \rref \cite{di2025dipole} would predict the bulk contribution $G_{\rm bulk}(\Delta)$ to be negligible with respect to the one coming from the FB, in this case we have to include this term in order to reproduce the correct shape of the interaction potential.
This stems from the fact that, under the considered choice of BCs, the system is mapped into a collection of 1D systems, entailing in particular the divergence of the density of states of bulk modes at the edge of each bands in the $\mu\not=0$ case.
Considering this term might prejudice the scaling of the interaction potential but, hopefully, this is not the case, as we show next.
The bulk modes' Green's function entering in the interaction potential reads
\begin{equation}
    G_{\rm bulk}(\Delta) = \sum_{\nu=\pm}\int_{-\pi}^\pi \mathrm{d} k \int_0^\pi \mathrm{d} q \frac{\langle a_{0m_i} | \mathcal{B}_\nu(k,q) \rangle\!\langle \mathcal{B}_\nu(k,q) | a_{0m_j} \rangle}{\Delta - \omega_\nu(k,q)}\,,
\end{equation}
which can be simplified to
\begin{equation}
    G_{\rm bulk}(\Delta) = \frac{2(\Delta+\mu)}{\pi^2}\int_{0}^\pi \mathrm{d} k \int_0^\pi \mathrm{d} q \frac{4\cos^2{(k/2)}\,\sin^2{(q)}\, \cos{\left(k|m|\right)}}{\left[\omega^2(k,q)-(\mu/J)^2\right]\left[\Delta^2-J^2\omega^2(k,q)\right]}\,.
\end{equation}
We now expand $[\Delta^2-J^2\omega^2(k,q)]^{-1}$ around its minimum as
\begin{equation}
    \frac{1}{\Delta^2 - J^2\omega^2(k,q)} \simeq \frac{1}{\Delta^2-\mu^2} \left[ 1 + \frac{J^2}{\Delta^2-\mu^2} \left( \frac{4-\beta^2}{4}(k-k_D)^2 + \beta^2(q-\pi)^2 \right) \right]
\end{equation}
which, upon substitution yields
\begin{equation}
    \begin{split}
    G_{\rm bulk}(\Delta) &\simeq \frac{8}{\Delta-\mu}\frac{1}{\pi^2} \int_{0}^\pi \mathrm{d} k\, \cos^2{(k/2)}\, \cos{\left(k|m|\right)}\left(1+J^2\frac{4-\beta^2 }{4 }\frac{(k-k_D)^2}{\Delta^2-\mu^2}\right) \left(\int_0^\pi \mathrm{d} q \frac{\sin^2{q}}{\omega^2(k,q)-(\mu/J)^2}\right)\\
    &+ \frac{4\beta^2}J^2{(\Delta-\mu)(\Delta^2-\mu^2)}\frac{1}{\pi^2} \int_{0}^\pi \mathrm{d} k\, \cos^2{(k/2)}\, \cos{\left(k|m|\right)} \left(\int_0^\pi \mathrm{d} q \frac{(q-\pi)^2\sin^2{q}}{\omega^2(k,q)-(\mu/J)^2}\right)\,.
    \end{split}
\end{equation}
The first $q$-integral in square brackets (first row) can be evaluated exactly near the Dirac point (which gives the maximal contribution to it), yielding
\begin{equation}
    \int_0^\pi \mathrm{d} q \frac{\sin^2{q}}{\omega^2(k_D},q)-(\mu/J)^2 = \frac{\pi}{2\beta^2}\,.
\end{equation}
Notice that only the term containing $(k-k_D)^2$ in the integrand will yield a power-law scaling, while the others will decay in a faster way. Hence, we now discard those other terms and focus only on the relevant one. We thus obtain
\begin{equation}
    G_{\rm bulk}(\Delta) = \frac{J^2(4-\beta^2)}{2\pi\beta^2(\Delta-\mu)(\Delta^2-\mu^2)} \int_0^\pi (1+\cos k)(k-k_D)^2 \cos(k|m|) \mathrm{d} k
\end{equation}
We now deal with the main integral. It is straightforward to prove, using integration by parts, that
\begin{equation}
    \int_0^\pi \mathrm{d} x (x-a)^2\cos{(mx)} = 2\frac{(\pi-a)(-1)^m+a}{m^2}\,.
\end{equation}
Thus, rewriting
\begin{equation}
    \int_0^\pi (1{+}\cos k)(k{-}k_D)^2 \cos(k|m|) \mathrm{d} k {=} \int_0^\pi (k{-}k_D)^2 \left[\cos(k|m|) {+} \frac{1}{2}\cos(k(|m|{+}1)) + \frac{1}{2}\cos(k(|m|{-}1))\right] \mathrm{d} k
\end{equation}
and applying the formula, we find that
\begin{equation}
    G_{\rm bulk} (\Delta) =\frac{4J^2}{\pi\beta^2} \frac{(4-\beta^2)\arccos{(\beta/2)}}{(\Delta-\mu)(\Delta^2-\mu^2)} \frac{1}{m^2} + o(m^{-2})\,.
\end{equation}
This result makes now clear that both contributions to the interaction potential $\mathcal{K}_{ij}$ have the same scaling for long distances, which is in particular the same as the cavity mode.
As a result, one gets power-law mediated interactions with an exponent equal to $-2$.

It is interesting exploring the relative magnitude of the two terms contributing to the interaction potential $\mathcal{K}_{ij}$, namely the dispersive bands contribution $G_{\rm bulk}(\Delta)$ and the FB contribution $G_{\rm FB}(\Delta) = (\Delta-\mu)^{-1} c(0,|m_i-m_j|)$, in the $m\gg 1$. Indeed, while the two terms have the same power-law scaling, there might be regimes in which one term is negligible compared to the other. In particular, if $G_{\rm bulk}(\Delta)\ll G_{\rm FB}(\Delta)$, the qubit-qubit interaction will be mediated solely by topologically-protected edge modes.
To the first non-trivial order, it is clear that
\begin{equation}
    \frac{G_{\rm bulk}(\Delta)}{G_{\rm FB}(\Delta)}\to \frac{2J^2}{\Delta^2 - \mu^2}\frac{k_D\sqrt{4-\beta^2}}{\beta}\frac{1}{\cos{(k_D m)}}\,,
\end{equation}
which can have either signs, depending on the sign of $\cos{(k_D m)}$ (notice that $\Delta^2-\mu^2$ is always negative).
We now notice the following. We can have $G_{\rm bulk}(\Delta)\ll G_{\rm FB}(\Delta)$ in two different ways: either $\mu\gg J$ and $\Delta \to 0$ (since $|\Delta^2-\mu^2|^{-1}$ has a minimum at $\Delta=0$ taking the value $\mu^2$), or $\beta \to 2^{-}$ (in which case $k_D\sqrt{4-\beta^2}/\beta \to 0)$. In these two parameter range, we will thus have interaction purely mediated by the edge states. On the other hand, if $\cos{(k_D m)}=0$, which can happen if (fixed $m$)
\begin{align}
    \beta = 2\cos{\left(\frac{(2p+1)\pi}{4m}\right)}\,, && p\in\mathbb{N}\,,
\end{align}
then $G_{\rm bulk}(\Delta)\gg G_{\rm FB}(\Delta)$ and the interaction will be mediated only by the dispersive modes.

\begin{figure}[h]
    \centering
    \includegraphics[width=0.8\linewidth]{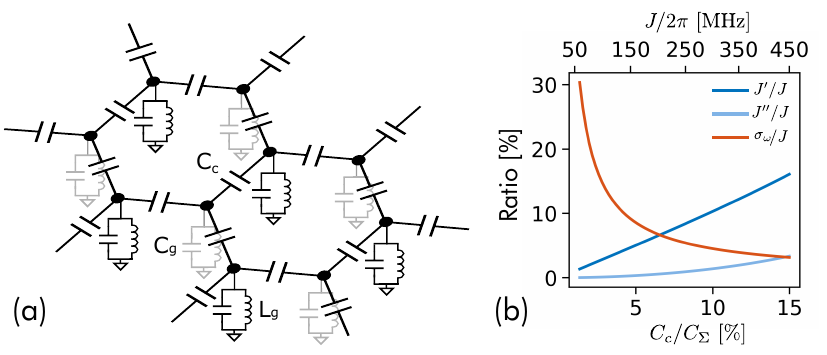}
    \caption{
    (a) Electrical circuit schematic of hexagonal lattice.
Each site contains an LC resonator ($L_g, C_g$) with neighboring sites coupled via capacitors $C_c$.
(b) Scaling of nearest-neighbor hopping $J$, second-nearest-neighbor hopping $J'$, third-nearest-neighbor hopping $J''$, and frequency disorder $\sigma_\omega$ with coupling strength $C_c/C_\Sigma$.
Stronger coupling enhances both the desired nearest-neighbor hopping $J$ and parasitic longer-range hoppings ($J'$, $J''$) while reducing relative disorder effects.}
    \label{fig:fig4}
\end{figure}

\section{ Experimental Implementation}
\label{sec:experiment}

The system presented in the main text can be directly implemented using superconducting lumped-element LC resonators \cite{underwood2016imaging, morvan2021observation, jouanny2025high} arranged in a honeycomb lattice [see \ref{fig:fig4}(a)].
Each lattice site consists of a parallel combination of inductor $L_g$ and capacitor $C_\mathrm{g}$ connected to ground.
Neighboring sites are coupled via capacitors $C_\mathrm{c}$, creating the nearest-neighbor hopping required for the tight-binding model.
The resonator frequency is determined by $\omega_r = 1/\sqrt{L_g C_\Sigma}$, where $C_\Sigma = C_\mathrm{g} + zC_\mathrm{c}$ represents the total capacitance at each site and $z$ is the coordination number.
In the tight-binding limit where $C_\mathrm{c} \ll C_\Sigma$, the hopping amplitude between neighboring sites takes the simple form $J = (C_\mathrm{c}/C_\Sigma) \omega_r/2$.
This capacitive coupling scheme works particularly well with compact, high-impedance resonators that have been implemented in circuit QED setups~\cite{scigliuzzo_controlling_2022, jouanny2025high}.
While we focus on capacitive coupling, inductive coupling provides an alternative implementation route with similar physics~\cite{youssefi2022topological, youssefi2025realization}.
To accurately model the system, we employ a full circuit analysis using the Lagrangian formalism, taking into account higher-order coupling terms that arise both from the inversion of the capacitance matrix and stray capacitance between next-nearest neighbor resonators~\cite{vool2017introduction, jouanny2025high}.
The system is described by constructing the complete capacitance matrix $\mathbf{C}$, which includes both $C_g$ and $C_\mathrm{c}$.
Combined with the inverse inductance matrix $\mathbf{L}^{-1}$ (diagonal with elements $1/L_g$), this yields the system Hamiltonian $\mathbf{H} = \sqrt{\mathbf{L}^{-1} \mathbf{C}^{-1}}$, whose eigenvalues correspond to the collective mode frequencies.
\begin{figure}[h]
    \centering
    \includegraphics[width=0.8\linewidth]{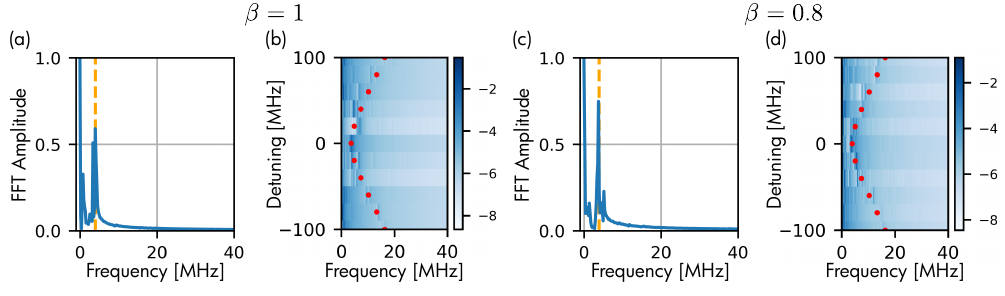}
    \caption{Single-qubit dynamics in a realistic setup for $\beta{=}1$ [panels (a)-(b)] and $\beta{=}0.8$ [panels (c)-(d)].
    The lattice is built using $\sim 460$ resonators. In both cases, we consider the presence of stray capacitances (${\sim}7.5\% \,C_c$), diagonal disorder (${\sim}0.5\%\, \omega_q$) and a parasitic qubit's decay rate ($\gamma_q/2\pi{\sim}100~\si{\kilo\hertz}$).
    The average hopping rate is $J/2\pi\sim 150 ~\si{\mega\hertz}$, while the coupling rate is either $g\sim 25 ~\si{\mega\hertz}$ ($\beta=1$) or $g\sim 35 ~\si{\mega\hertz}$ ($\beta=0.8$). Using these parameters, we have $J'/J\simeq 11.4\%$.
    The qubit is prepared in the excited state and then evolves for $5~\si{\micro\second}$ with a sampling every $4~\si{\nano\second}$.
    (a,c) Normalized amplitude of the Fast Fourier Transform (FFT) of the qubit's population.
    The orange dashed line corresponds to the predicted Rabi frequency $\Omega$.
    Panels (b),(d)  Logarithm of the FFT's amplitude of the qubit's population for different values of detuning ($\Delta=0$ corresponds to resonance with the FB).
    Red points correspond to the predicted Rabi frequency for each value of detuning.
    }
    \label{fig:disorder_1qubit}
\end{figure}

Several non-idealities arise in circuit implementations.
First, due to the finite size of the device, resonators at the edges have a different coordination number ($z = 2$) than those in the bulk ($z = 3$), resulting in a frequency shift for edge resonators.
Hence, for sites with different coordination numbers, the ground capacitance is adjusted to maintain a fixed $C_\Sigma$, thus ensuring uniform on-site frequencies across the lattice.~\cite{scigliuzzo_controlling_2022,jouanny2025high}
Second, long-range hopping terms arising from two sources: direct stray capacitance between next-nearest neighbor resonators and the intrinsic inversion of the capacitance matrix.
Even when only nearest-neighbor capacitive couplings \( C_\mathrm{c} \) are physically implemented, the circuit equations generate second-nearest-neighbor hoppings \( J' \), third-nearest-neighbor hoppings \( J'' \), and higher-order terms.
Third, fabrication variations in both inductance and capacitance values lead to disorder in circuit parameters.
While disorder in both on-site frequencies and hopping amplitudes is present, only the on-site frequency disorder (diagonal disorder) breaks the chiral symmetry of the system.
Disorder in hopping amplitudes preserves chiral symmetry and thus does not affect the frequency of the edge modes discussed in the main text~\cite{ramachandran_chiral_2017}.
Therefore, our analysis focuses on the on-site frequency disorder with standard deviation \( \sigma_\omega \), arising from variations in \( L_g \) and \( C_\mathrm{g} \).
\begin{figure}[h]
    \centering
    \includegraphics[width=0.8\linewidth]{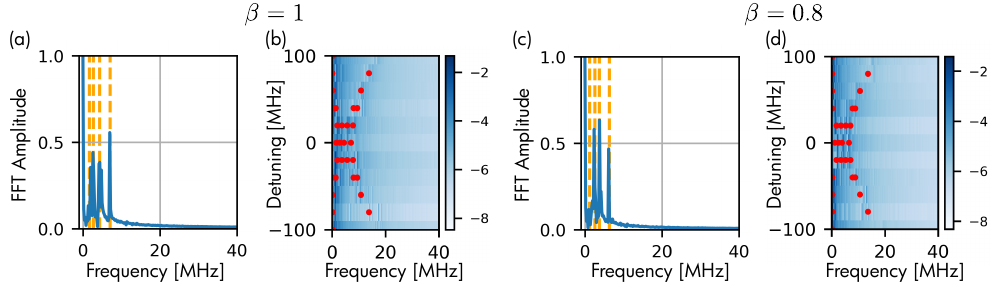}
    \caption{Two-qubit dynamics in a realistic setup for $\beta{=}1$ [panel (a-b)] and $\beta{=}0.8$ [panel (c-d)].
    The lattice is built using $\sim 460$ resonators. In both cases, we consider the presence of stray capacitances (${\sim}7.5\% \,C_c$), diagonal disorder (${\sim}0.5\%\, \omega_q$) and a parasitic qubit's decay rate ($\gamma_q/2\pi{\sim}100~\si{\kilo\hertz}$).
    The average hopping rate is $J/2\pi\sim 150 ~\si{\mega\hertz}$, while the coupling rate is $g\sim 35 ~\si{\mega\hertz}$. Using these parameters, we have $J'/J\simeq 11.4\%$.
    The first (second) qubit is prepared in the excited (ground) state and then the system evolves for $5~\si{\micro\second}$ with a sampling every $4~\si{\nano\second}$.
    (a,c) Normalized amplitude of the Fast Fourier Transform (FFT) of the first qubit's population on resonance with the FB.
    The orange dashed line corresponds to the four predicted frequencies.
    (b,d) Logarithm of the FFT's amplitude of the first qubit's population for different values of detuning ($\Delta=0$ being on resonance with the FB).
    Red points correspond to the predicted Rabi frequency for each value of detuning.} 
    \label{fig:disorder_2qubit}
\end{figure}

In each case, chiral-symmetry breaking gives the zero-energy flat band a finite spectral width and causes the edge modes, which ideally reside only on the $a$-sublattice, to acquire non-zero amplitude on both sublattices.
Preliminary analyses show that, in order to observe the presented phenomenology, we need to work in a parameter region where this spectral width is small compared to the qubit coupling $g/2\pi$ [see \ref{fig:disorder_1qubit} and \ref{fig:disorder_2qubit}].
Moreover, for sufficiently large disorder compared to the hopping strength, the system enters an Anderson localization regime \cite{anderson_absence_1958} where all modes become spatially localized at random positions.
In light of the above, the choice of circuit parameters involves a fundamental trade-off [see \ref{fig:fig4}(b)].
For fixed resonator frequency, increasing the coupling strength by raising the ratio $C_\mathrm{c}/C_\Sigma$ improves robustness against disorder by increasing all hopping strengths ($J$, $J'$, $J''$) but maintaining fixed frequency disorder.
However, stronger coupling simultaneously enhances the unwanted higher-order hopping terms, making the nearest-neighbor approximation in the main text less accurate.
Taking into account the effects mentioned above, we perform numerical simulations of the dynamics of a single [\ref{fig:disorder_1qubit}] and two qubits [\ref{fig:disorder_2qubit}] coupled to the zigzag edge (parameters of the simulations are in the captions).
In this way, we find that the parameters mentioned in the main text -- i.e., $\omega_r/2\pi = 6$ GHz, $J/2\pi = 100$-200 MHz and $g/2\pi = 20$ MHz -- represent a reasonable compromise (see \ref{fig:fig4}b).
Moreover, for these values, the next-nearest-neighbor hopping $J'/J$ and frequency disorder $\sigma_\omega/J$ are in the range $5-10\,\%$ with the spectral width of the flat band being reasonably smaller than $g/2\pi$.
This analysis shows that a proof--of--principle demonstration of the physics discussed in this work, namely the coherent dynamics induced by the presence of edge modes, is within reach of circuit--QED platforms.
\printbibliography

@article{ozawa_topological_2019,
  title = {Topological photonics},
  author = {Ozawa, Tomoki and Price, Hannah M. and Amo, Alberto and Goldman, Nathan and Hafezi, Mohammad and Lu, Ling and Rechtsman, Mikael C. and Schuster, David and Simon, Jonathan and Zilberberg, Oded and Carusotto, Iacopo},
  journal = {Rev. Mod. Phys.},
  volume = {91},
  issue = {1},
  pages = {015006},
  numpages = {76},
  year = {2019},
  month = {Mar},
  publisher = {American Physical Society},
  doi = {10.1103/RevModPhys.91.015006},
  url = {https://link.aps.org/doi/10.1103/RevModPhys.91.015006}
}

@article{diekmann_2026_enabling,
      title={Enabling Deterministic Passive Quantum State Transfer with Giant Atoms}, 
      author={Oliver Diekmann and Enrico Di Benedetto and Nicolas Jungwirth and Daniele De Bernardis and Zeyu Kuang and Francesco Ciccarello and Stefan Rotter and Peter Rabl and Alejandro Gonz{\'a}lez-Tudela and Carlos Gonz{\'a}lez-Ballestero},
      journal={arXiv preprint arXiv:2605.12018},
      year={2026},
      url={https://arxiv.org/abs/2605.12018},
}

@article{fink_dressed_2009,
  title = {Dressed Collective Qubit States and the Tavis-Cummings Model in Circuit QED},
  author = {Fink, J. M. and Bianchetti, R. and Baur, M. and Goppl, M. and Steffen, L. and Filipp, S. and Leek, P. J. and Blais, A. and Wallraff, A.},
  journal = {Phys. Rev. Lett.},
  volume = {103},
  issue = {8},
  pages = {083601},
  numpages = {4},
  year = {2009},
  month = {Aug},
  publisher = {American Physical Society},
  doi = {10.1103/PhysRevLett.103.083601},
  url = {https://link.aps.org/doi/10.1103/PhysRevLett.103.083601}
}

@article{tavis_exact_1968,
  title = {Exact Solution for an $N$-Molecule---Radiation-Field Hamiltonian},
  author = {Tavis, Michael and Cummings, Frederick W.},
  journal = {Phys. Rev.},
  volume = {170},
  issue = {2},
  pages = {379--384},
  numpages = {0},
  year = {1968},
  month = {Jun},
  publisher = {American Physical Society},
  doi = {10.1103/PhysRev.170.379},
  url = {https://link.aps.org/doi/10.1103/PhysRev.170.379}
}

@article{gelhausen2018dissipative,
  title = {Dissipative Dicke model with collective atomic decay: Bistability, noise-driven activation, and the nonthermal first-order superradiance transition},
  author = {Gelhausen, Jan and Buchhold, Michael},
  journal = {Phys. Rev. A},
  volume = {97},
  issue = {2},
  pages = {023807},
  numpages = {13},
  year = {2018},
  month = {Feb},
  publisher = {American Physical Society},
  doi = {10.1103/PhysRevA.97.023807},
  url = {https://link.aps.org/doi/10.1103/PhysRevA.97.023807}
}

@article{o2025circuit,
  title={A Circuit-QED Lattice System with Flexible Connectivity and Gapped Flat Bands for Photon-Mediated Spin Models},
  author={O'Brien, Kellen and Amouzegar, Maya and Lee, Won Chan and Ritter, Martin and Koll{\'a}r, Alicia J},
  journal={arXiv:2505.05559},
  year={2025},
  url={https://arxiv.org/abs/2505.05559}
}

@article{kim2020recent,
  title={Recent advances in 2D, 3D and higher-order topological photonics},
  author={Kim, Minkyung and Jacob, Zubin and Rho, Junsuk},
  journal={Light: Science \& Applications},
  volume={9},
  number={1},
  pages={130},
  year={2020},
  publisher={Nature Publishing Group UK London},
  url={https://www.nature.com/articles/s41377-020-0331-y}
}

@article{kim2021quantum,
  title={Quantum electrodynamics in a topological waveguide},
  author={Kim, Eunjong and Zhang, Xueyue and Ferreira, Vinicius S and Banker, Jash and Iverson, Joseph K and Sipahigil, Alp and Bello, Miguel and Gonz{\'a}lez-Tudela, Alejandro and Mirhosseini, Mohammad and Painter, Oskar},
  journal={Physical Review X},
  volume={11},
  number={1},
  pages={011015},
  year={2021},
  publisher={APS},
  url={https://journals.aps.org/prx/abstract/10.1103/PhysRevX.11.011015}
}

@article{bello2019unconventional,
    author = {M. Bello  and G. Platero  and J. I. Cirac  and A. Gonz{\'a}lez-Tudela },
    title = {Unconventional quantum optics in topological waveguide QED},
    journal = {Science Advances},
    volume = {5},
    number = {7},
    pages = {eaaw0297},
    year = {2019},
    doi = {10.1126/sciadv.aaw0297},
    URL = {https://www.science.org/doi/abs/10.1126/sciadv.aaw0297}
}

@article{ferreira2021collapse,
  title = {Collapse and Revival of an Artificial Atom Coupled to a Structured Photonic Reservoir},
  author = {Ferreira, Vinicius S. and Banker, Jash and Sipahigil, Alp and Matheny, Matthew H. and Keller, Andrew J. and Kim, Eunjong and Mirhosseini, Mohammad and Painter, Oskar},
  journal = {Phys. Rev. X},
  volume = {11},
  issue = {4},
  pages = {041043},
  numpages = {27},
  year = {2021},
  month = {Dec},
  publisher = {American Physical Society},
  doi = {10.1103/PhysRevX.11.041043},
  url = {https://link.aps.org/doi/10.1103/PhysRevX.11.041043}
}

@article{jouanny2025superstrong,
  title={Superstrong Dynamics and Directional Emission of a Giant Atom in a Structured Bath},
  author={Jouanny, Vincent and Peyruchat, L{\'e}o and Scigliuzzo, Marco and Mercurio, Alberto and Di Benedetto, Enrico and De Bernardis, Daniele and Sbroggi{\`o}, Davide and Frasca, Simone and Savona, Vincenzo and Ciccarello, Francesco and others},
  journal={arXiv preprint arXiv:2509.01579},
  url = {https://doi.org/10.48550/arXiv.2509.01579},
  year={2025}
}

@article{pakkiam2024experimental,
  title={Experimental realization of qubit-state-controlled directional edge states in waveguide QED},
  author={Pakkiam, Prasanna and Kumar, N Pradeep and Chiu, Chun-Ching and Sommers, David and Pletyukhov, Mikhail and Fedorov, Arkady},
  journal={arXiv preprint arXiv:2411.05271},
url = {https://doi.org/10.48550/arXiv.2411.05271},
  year={2024}
}

@article{gong2016topological,
  title={Topological phases with long-range interactions},
  author={Gong, Z-X and Maghrebi, Mohammad F and Hu, Anzi and Wall, Michael L and Foss-Feig, Michael and Gorshkov, Alexey V},
  journal={Physical Review B},
  volume={93},
  number={4},
  pages={041102},
  year={2016},
  publisher={APS},
  url={https://journals.aps.org/prb/abstract/10.1103/PhysRevB.93.041102}
}

@article{islam2013emergence,
  title={Emergence and frustration of magnetism with variable-range interactions in a quantum simulator},
  author={Islam, R and Senko, Crystal and Campbell, Wes C and Korenblit, S and Smith, J and Lee, A and Edwards, EE and Wang, C-CJ and Freericks, JK and Monroe, C},
  journal={science},
  volume={340},
  number={6132},
  pages={583--587},
  year={2013},
  publisher={American Association for the Advancement of Science},
  url={https://www.science.org/doi/10.1126/science.1232296}
}

@article{vzunkovivc2018dynamical,
  title={Dynamical quantum phase transitions in spin chains with long-range interactions: Merging different concepts of nonequilibrium criticality},
  author={{\v{Z}}unkovi{\v{c}}, Bojan and Heyl, Markus and Knap, Michael and Silva, Alessandro},
  journal={Physical review letters},
  volume={120},
  number={13},
  pages={130601},
  year={2018},
  publisher={APS},
  url={https://journals.aps.org/prl/abstract/10.1103/PhysRevLett.120.130601}
}

@article{gong2016kaleidoscope,
  title={Kaleidoscope of quantum phases in a long-range interacting spin-1 chain},
  author={Gong, Z-X and Maghrebi, Mohammad F and Hu, Anzi and Foss-Feig, Michael and Richerme, Phillip and Monroe, Christopher and Gorshkov, Alexey V},
  journal={Physical Review B},
  volume={93},
  number={20},
  pages={205115},
  year={2016},
  publisher={APS},
  url={https://journals.aps.org/prb/abstract/10.1103/PhysRevB.93.205115}
}

@article{Sundaresan2019,
author = {Sundaresan, Neereja M. and Lundgren, Rex and Zhu, Guanyu and Gorshkov, Alexey V. and Houck, Andrew A.},
doi = {10.1103/PhysRevX.9.011021},
issn = {21603308},
journal = {Physical Review X},
month = {feb},
number = {1},
pages = {011021},
title = {{Interacting Qubit-Photon Bound States with Superconducting Circuits}},
url = {https://link.aps.org/doi/10.1103/PhysRevX.9.011021},
volume = {9},
year = {2019}
}

@article{fujita1996peculiar,
  title={Peculiar localized state at zigzag graphite edge},
  author={Fujita, Mitsutaka and Wakabayashi, Katsunori and Nakada, Kyoko and Kusakabe, Koichi},
  journal={Journal of the Physical Society of Japan},
  volume={65},
  number={7},
  pages={1920--1923},
  year={1996},
  publisher={The Physical Society of Japan},
  url={https://journals.jps.jp/doi/10.1143/JPSJ.65.1920}
}

@article{nakada1996edge,
  title={Edge state in graphene ribbons: Nanometer size effect and edge shape dependence},
  author={Nakada, Kyoko and Fujita, Mitsutaka and Dresselhaus, Gene and Dresselhaus, Mildred S},
  journal={Physical Review B},
  volume={54},
  number={24},
  pages={17954},
  year={1996},
  publisher={APS},
  url={https://journals.aps.org/prb/abstract/10.1103/PhysRevB.54.17954}
}

@article{yao2013topologically,
  title={Topologically protected quantum state transfer in a chiral spin liquid},
  author={Yao, Norman Y and Laumann, Chris R and Gorshkov, Alexey V and Weimer, Hendrik and Jiang, Liang and Cirac, J Ignacio and Zoller, Peter and Lukin, Mikhail D},
  journal={Nature communications},
  volume={4},
  number={1},
  pages={1585},
  year={2013},
  publisher={Nature Publishing Group UK London},
  url={https://www.nature.com/articles/ncomms2531}
}

@article{dlaska2017robust,
  title={Robust quantum state transfer via topologically protected edge channels in dipolar arrays},
  author={Dlaska, Clemens and Vermersch, Beno{\^\i}t and Zoller, Peter},
  journal={Quantum Science and Technology},
  volume={2},
  number={1},
  pages={015001},
  year={2017},
  publisher={IOP Publishing},
  url={https://iopscience.iop.org/article/10.1088/2058-9565/2/1/015001}
}

@article{bouscal2024systematic,
  title={Systematic design of a robust half-W1 photonic crystal waveguide for interfacing slow light and trapped cold atoms},
  author={Bouscal, Adrien and Kemiche, Malik and Mahapatra, Sukanya and Fayard, Nikos and Berroir, J{\'e}r{\'e}my and Ray, Tridib and Greffet, Jean-Jacques and Raineri, Fabrice and Levenson, Ariel and Bencheikh, Kamel and others},
  journal={New Journal of Physics},
  volume={26},
  number={2},
  pages={023026},
  year={2024},
  publisher={IOP Publishing},
  url={https://iopscience.iop.org/article/10.1088/1367-2630/ad23a4}
}

@article{tan_edge_2021,
  title = {Edge state in AB-stacked bilayer graphene and its correspondence with the Su-Schrieffer-Heeger ladder},
  author = {Tan, Tixuan and Li, Ci and Yao, Wang},
  journal = {Phys. Rev. B},
  volume = {104},
  issue = {24},
  pages = {245419},
  numpages = {14},
  year = {2021},
  month = {Dec},
  publisher = {American Physical Society},
  doi = {10.1103/PhysRevB.104.245419},
  url = {https://link.aps.org/doi/10.1103/PhysRevB.104.245419}
}

@article{kohmoto2007zero,
  title={Zero modes and edge states of the honeycomb lattice},
  author={Kohmoto, Mahito and Hasegawa, Yasumasa},
  journal={Physical Review B—Condensed Matter and Materials Physics},
  volume={76},
  number={20},
  pages={205402},
  year={2007},
  publisher={APS},
  url = {https://link.aps.org/doi/10.1103/PhysRevB.76.205402}
}

@article{Castro_Graphene_2009,
  title = {The electronic properties of graphene},
  author = {Castro Neto, A. H. and Guinea, F. and Peres, N. M. R. and Novoselov, K. S. and Geim, A. K.},
  journal = {Rev. Mod. Phys.},
  volume = {81},
  issue = {1},
  pages = {109--162},
  numpages = {0},
  year = {2009},
  month = {Jan},
  publisher = {American Physical Society},
  doi = {10.1103/RevModPhys.81.109},
  url = {https://link.aps.org/doi/10.1103/RevModPhys.81.109}
}

@article{owens2022chiral,
  title={Chiral cavity quantum electrodynamics},
  author={Owens, John Clai and Panetta, Margaret G and Saxberg, Brendan and Roberts, Gabrielle and Chakram, Srivatsan and Ma, Ruichao and Vrajitoarea, Andrei and Simon, Jonathan and Schuster, David I},
  journal={Nature Physics},
  volume={18},
  number={9},
  pages={1048--1052},
  year={2022},
  publisher={Nature Publishing Group UK London},
  url={https://www.nature.com/articles/s41567-022-01671-3}
}

@article{di2025dipole,
  title={Dipole-dipole interactions mediated by a photonic flat band},
  author={Di Benedetto, Enrico and Gonz{\'a}lez-Tudela, Alejandro and Ciccarello, Francesco},
  journal={Quantum},
  volume={9},
  pages={1671},
  year={2025},
  publisher={Verein zur F{\"o}rderung des Open Access Publizierens in den Quantenwissenschaften},
  url={https://quantum-journal.org/papers/q-2025-03-25-1671/}
}

@article{lambropoulos_fundamental_2000,
	title = {Fundamental quantum optics in structured reservoirs},
	volume = {63},
	issn = {0034-4885},
	url = {https://dx.doi.org/10.1088/0034-4885/63/4/201},
	doi = {10.1088/0034-4885/63/4/201},
	language = {english},
	number = {4},
	urldate = {2024-02-20},
	journal = {Rep. Prog. Phys.},
	author = {Lambropoulos, P. and Nikolopoulos, Georgios M. and Nielsen, Torben R. and Bay, Søren},
	month = apr,
	year = {2000},
	pages = {455},
}

@incollection{economou_greens_1979,
	address = {Berlin, Heidelberg},
	series = {Springer {Series} in {Solid}-{State} {Sciences}},
	title = {Green’s {Functions} for {Tight} {Binding} {Hamiltonians}},
	isbn = {978-3-662-11900-6},
	url = {https://doi.org/10.1007/978-3-662-11900-6_5},
	language = {english},
	urldate = {2024-02-20},
	booktitle = {Green’s {Functions} in {Quantum} {Physics}},
	publisher = {Springer},
	author = {Economou, Eleftherios N.},
	editor = {Economou, Eleftherios N.},
	year = {1979},
	doi = {10.1007/978-3-662-11900-6_5},
	pages = {71--91},
}

@book{asboth2016short,
  title={A short course on topological insulators},
  author={Asb{\'o}th, J{\'a}nos K and Oroszl{\'a}ny, L{\'a}szl{\'o} and P{\'a}lyi, Andr{\'a}s},
  volume={919},
  year={2016},
  publisher={Springer},
  url={https://link.springer.com/book/10.1007/978-3-319-25607-8}
}

@article{vega_topological_2023,
  title = {Topological multimode waveguide QED},
  author = {Vega, C. and Porras, D. and Gonz{\'a}lez-Tudela, A.},
  journal = {Phys. Rev. Res.},
  volume = {5},
  issue = {2},
  pages = {023031},
  numpages = {19},
  year = {2023},
  month = {Apr},
  publisher = {American Physical Society},
  doi = {10.1103/PhysRevResearch.5.023031},
  url = {https://link.aps.org/doi/10.1103/PhysRevResearch.5.023031}
}

@article{ryu2002topological,
  title={Topological origin of zero-energy edge states in particle-hole symmetric systems},
  author={Ryu, Shinsei and Hatsugai, Yasuhiro},
  journal={Physical review letters},
  volume={89},
  number={7},
  pages={077002},
  year={2002},
  publisher={APS},
  url={https://journals.aps.org/prl/abstract/10.1103/PhysRevLett.89.077002}
}

@article{windt_retardation_2025,
  title = {Effects of Retardation on Many-Body Superradiance in Chiral Waveguide QED},
  author = {Windt, Bennet and Bello, Miguel and Malz, Daniel and Cirac, J. Ignacio},
  journal = {Phys. Rev. Lett.},
  volume = {134},
  issue = {17},
  pages = {173601},
  numpages = {8},
  year = {2025},
  month = {Apr},
  publisher = {American Physical Society},
  doi = {10.1103/PhysRevLett.134.173601},
  url = {https://link.aps.org/doi/10.1103/PhysRevLett.134.173601}
}

@article{rhim_classification_2019,
	title = {Classification of flat bands according to the band-crossing singularity of {Bloch} wave functions},
	volume = {99},
	url = {https://link.aps.org/doi/10.1103/PhysRevB.99.045107},
	doi = {10.1103/PhysRevB.99.045107},
	number = {4},
	urldate = {2024-02-20},
	journal = {Phys. Rev. B},
	author = {Rhim, Jun-Won and Yang, Bohm-Jung},
	month = jan,
	year = {2019},
	
	pages = {045107},
}

@article{ramachandran_chiral_2017,
	title = {Chiral flat bands: {Existence}, engineering, and stability},
	volume = {96},
	shorttitle = {Chiral flat bands},
	url = {https://link.aps.org/doi/10.1103/PhysRevB.96.161104},
	doi = {10.1103/PhysRevB.96.161104},
	number = {16},
	urldate = {2024-02-20},
	journal = {Phys. Rev. B},
	author = {Ramachandran, Ajith and Andreanov, Alexei and Flach, Sergej},
	month = oct,
	year = {2017},
	
	pages = {161104},
}

@article{lorenzo_quantum_2017,
	title = {Quantum non-{Markovianity} induced by {Anderson} localization},
	volume = {7},
	copyright = {2017 The Author(s)},
	issn = {2045-2322},
	url = {https://www.nature.com/articles/srep42729},
	doi = {10.1038/srep42729},
	language = {english},
	number = {1},
	urldate = {2024-02-20},
	journal = {Sci Rep},
	author = {Lorenzo, Salvatore and Lombardo, Federico and Ciccarello, Francesco and Palma, G. Massimo},
	month = feb,
	year = {2017},
	pages = {42729},
}

@article{anderson_absence_1958,
	title = {Absence of {Diffusion} in {Certain} {Random} {Lattices}},
	volume = {109},
	url = {https://link.aps.org/doi/10.1103/PhysRev.109.1492},
	doi = {10.1103/PhysRev.109.1492},
	number = {5},
	urldate = {2024-02-20},
	journal = {Phys. Rev.},
	author = {Anderson, P. W.},
	month = mar,
	year = {1958},
	
	pages = {1492--1505},
}

@article{park2024quasi,
  title={Quasi-localization and Wannier obstruction in partially flat bands},
  author={Park, Jin-Hong and Rhim, Jun-Won},
  journal={Communications Physics},
  volume={7},
  number={1},
  pages={179},
  year={2024},
  publisher={Nature Publishing Group UK London},
  url={https://www.nature.com/articles/s42005-024-01679-6}
}

@article{suarez2025chiral,
  title={Chiral quantum optics: recent developments and future directions},
  author={Su{\'a}rez-Forero, DG and Jalali Mehrabad, M and Vega, C and Gonz{\'a}lez-Tudela, A and Hafezi, M},
  journal={Prx Quantum},
  volume={6},
  number={2},
  pages={020101},
  year={2025},
  publisher={APS},
  url={https://journals.aps.org/prxquantum/abstract/10.1103/PRXQuantum.6.020101}
}

@article{pereira_graphene_2006,
  title = {Disorder Induced Localized States in Graphene},
  author = {Pereira, Vitor M. and Guinea, F. and Lopes dos Santos, J. M. B. and Peres, N. M. R. and Castro Neto, A. H.},
  journal = {Phys. Rev. Lett.},
  volume = {96},
  issue = {3},
  pages = {036801},
  numpages = {4},
  year = {2006},
  month = {Jan},
  publisher = {American Physical Society},
  doi = {10.1103/PhysRevLett.96.036801},
  url = {https://link.aps.org/doi/10.1103/PhysRevLett.96.036801}
}

@article{leykam_artificial_2018,
	title = {Artificial flat band systems: from lattice models to experiments},
	volume = {3},
	issn = {null},
	shorttitle = {Artificial flat band systems},
	url = {https://doi.org/10.1080/23746149.2018.1473052},
	doi = {10.1080/23746149.2018.1473052},
	number = {1},
	urldate = {2024-02-20},
	journal = {Advances in Physics: X},
	author = {Leykam, Daniel and Andreanov, Alexei and Flach, Sergej},
	month = jan,
	year = {2018},
	pages = {1473052},
}

@article{redondo2021quantum,
  title={Quantum electrodynamics in anisotropic and tilted Dirac photonic lattices},
  author={Redondo-Yuste, Jaime and de Paz, Mar{\'\i}a Blanco and Huidobro, Paloma A and Gonz{\'a}lez-Tudela, Alejandro},
  journal={New Journal of Physics},
  volume={23},
  number={10},
  pages={103018},
  year={2021},
  publisher={IOP Publishing},
  url={https://iopscience.iop.org/article/10.1088/1367-2630/ac27e0}
}

@article{gonzalez2018exotic,
  title={Exotic quantum dynamics and purely long-range coherent interactions in Dirac conelike baths},
  author={Gonz{\'a}lez-Tudela, Alejandro and Cirac, J Ignacio},
  journal={Physical Review A},
  volume={97},
  number={4},
  pages={043831},
  year={2018},
  publisher={APS},
  url={https://journals.aps.org/pra/abstract/10.1103/PhysRevA.97.043831}
}

@article{chen_impossibility_2014,
	title = {The impossibility of exactly flat non-trivial {Chern} bands in strictly local periodic tight binding models},
	volume = {47},
	issn = {1751-8121},
	url = {https://dx.doi.org/10.1088/1751-8113/47/15/152001},
	doi = {10.1088/1751-8113/47/15/152001},
	language = {english},
	number = {15},
	urldate = {2024-02-22},
	journal = {J. Phys. A: Math. Theor.},
	author = {Chen, Li and Mazaheri, Tahereh and Seidel, Alexander and Tang, Xiang},
	month = mar,
	year = {2014},
	
	pages = {152001},
}

@article{rhim_singular_2021,
	title = {Singular flat bands},
	volume = {6},
	issn = {null},
	url = {https://doi.org/10.1080/23746149.2021.1901606},
	doi = {10.1080/23746149.2021.1901606},
	number = {1},
	urldate = {2024-02-22},
	journal = {Advances in Physics: X},
	author = {Rhim, Jun-Won and Yang, Bohm-Jung},
	month = jan,
	year = {2021},
	pages = {1901606},
}

@article{weber20242024,
  title={2024 roadmap on 2D topological insulators},
  author={Weber, Bent and Fuhrer, Michael S and Sheng, Xian-Lei and Yang, Shengyuan A and Thomale, Ronny and Shamim, Saquib and Molenkamp, Laurens W and Cobden, David and Pesin, Dmytro and Zandvliet, Harold JW and others},
  journal={Journal of Physics: Materials},
  volume={7},
  number={2},
  pages={022501},
  year={2024},
  publisher={IOP Publishing},
  url={https://iopscience.iop.org/article/10.1088/2515-7639/ad2083}
}

@article{bose_quantum_2003,
  title = {Quantum Communication through an Unmodulated Spin Chain},
  author = {Bose, Sougato},
  journal = {Phys. Rev. Lett.},
  volume = {91},
  issue = {20},
  pages = {207901},
  numpages = {4},
  year = {2003},
  month = {Nov},
  publisher = {American Physical Society},
  doi = {10.1103/PhysRevLett.91.207901},
  url = {https://link.aps.org/doi/10.1103/PhysRevLett.91.207901}
}

@article{douglas_quantum_2015,
	title = {Quantum many-body models with cold atoms coupled to photonic crystals},
	volume = {9},
	copyright = {2015 Springer Nature Limited},
	issn = {1749-4893},
	url = {https://www.nature.com/articles/nphoton.2015.57},
	doi = {10.1038/nphoton.2015.57},
	language = {english},
	number = {5},
	urldate = {2024-02-20},
	journal = {Nature Photon},
	author = {Douglas, J. S. and Habibian, H. and Hung, C.-L. and Gorshkov, A. V. and Kimble, H. J. and Chang, D. E.},
	month = may,
	year = {2015},
	pages = {326--331},
}

@article{gonzalez-tudela_subwavelength_2015,
	title = {Subwavelength vacuum lattices and atom–atom interactions in two-dimensional photonic crystals},
	volume = {9},
	copyright = {2015 Springer Nature Limited},
	issn = {1749-4893},
	url = {https://www.nature.com/articles/nphoton.2015.54},
	doi = {10.1038/nphoton.2015.54},
	language = {english},
	number = {5},
	urldate = {2024-02-20},
	journal = {Nature Photon},
	author = {Gonz{\'a}lez-Tudela, A. and Hung, C.-L. and Chang, D. E. and Cirac, J. I. and Kimble, H. J.},
	month = may,
	year = {2015},
	pages = {320--325},
}

@article{LeonforteVDS,
	title = {Vacancy-like Dressed States in Topological Waveguide QED},
	author = {Leonforte, Luca and Carollo, Angelo and Ciccarello, Francesco},
	journal = {Phys. Rev. Lett.},
	volume = {126},
	issue = {6},
	pages = {063601},
	numpages = {7},
	year = {2021},
	month = {Feb},
	publisher = {American Physical Society},
	doi = {10.1103/PhysRevLett.126.063601},
	url = {https://link.aps.org/doi/10.1103/PhysRevLett.126.063601}
}

@inproceedings{Chochon:25,
    author = {Ana{i}s Chochon and Adrien Bouscal and Sukanya Mahapatra and Idriss Douss and Val\`{e}re Sautel and Malik Kemiche and Nikos Fayard and J\'{e}r\'{e}my Berroir and Tridib Ray and Jean-Jacques Greffet and Fabrice Raineri and Ariel Levenson and Kamel Bencheikh and Christophe Sauvan and Alban Urvoy and Julien Laurat},
    booktitle = {Optica Quantum 2.0 Conference and Exhibition},
    journal = {Optica Quantum 2.0 Conference and Exhibition},
    pages = {QTh4A.7},
    publisher = {Optica Publishing Group},
    title = {Coupling slow-mode nanophotonics and cold atoms: a versatile Waveguide QED platform},
    year = {2025},
    url = {https://opg.optica.org/abstract.cfm?URI=QUANTUM-2025-QTh4A.7},
    doi = {10.1364/QUANTUM.2025.QTh4A.7},
}

@article{jamadi_2021_optically,
      title={Optically defined cavities in driven-dissipative photonic lattices}, 
      author={O. Jamadi and B. Real and K. Sawicki and C. Hainaut and A. Gonzalez-Tudela and N. Pernet and I. Sagnes and M. Morassi and A. Lemaitre and L. Le Gratiet and A. Harouri and S. Ravets and J. Bloch and A. Amo},
      year={2021},
      journal={arXiv preprint arXiv:2112.07753},
      url={https://arxiv.org/abs/2112.07753},
}

@article{leonforte_quantum_2024,
doi = {10.1088/2058-9565/ada08d},
url = {https://dx.doi.org/10.1088/2058-9565/ada08d},
year = {2024},
month = {dec},
publisher = {IOP Publishing},
volume = {10},
number = {1},
pages = {015057},
author = {Leonforte, Luca and Sun, Xuejian and Valenti, Davide and Spagnolo, Bernardo and Illuminati, Fabrizio and Carollo, Angelo and Ciccarello, Francesco},
title = {Quantum optics with giant atoms in a structured photonic bath},
journal = {Quantum Science and Technology}
}

@article{ritsch_cold_2013,
  title = {Cold atoms in cavity-generated dynamical optical potentials},
  author = {Ritsch, Helmut and Domokos, Peter and Brennecke, Ferdinand and Esslinger, Tilman},
  journal = {Rev. Mod. Phys.},
  volume = {85},
  issue = {2},
  pages = {553--601},
  numpages = {0},
  year = {2013},
  month = {Apr},
  publisher = {American Physical Society},
  doi = {10.1103/RevModPhys.85.553},
  url = {https://link.aps.org/doi/10.1103/RevModPhys.85.553}
}

@article{rechtsman2013topological,
  title = {Topological Creation and Destruction of Edge States in Photonic Graphene},
  author = {Rechtsman, Mikael C. and Plotnik, Yonatan and Zeuner, Julia M. and Song, Daohong and Chen, Zhigang and Szameit, Alexander and Segev, Mordechai},
  journal = {Phys. Rev. Lett.},
  volume = {111},
  issue = {10},
  pages = {103901},
  numpages = {5},
  year = {2013},
  month = {Sep},
  publisher = {American Physical Society},
  doi = {10.1103/PhysRevLett.111.103901},
  url = {https://link.aps.org/doi/10.1103/PhysRevLett.111.103901}
}

@book{breuer2002theory,
  title={The theory of open quantum systems},
  author={Breuer, Heinz-Peter and Petruccione, Francesco},
  year={2002},
  publisher={OUP Oxford},
url={https://global.oup.com/academic/product/the-theory-of-open-quantum-systems-9780198520634?cc=de&lang=en&}
}

@article{kim2025real,
  title={Real space decay of flat band projectors from compact localized states},
  author={Kim, Yeongjun and Flach, Sergej and Andreanov, Alexei},
  journal={arXiv preprint arXiv:2510.17258},
  year={2025},
  url = {https://arxiv.org/abs/2510.17258}
}

@article{zhang2023superconducting,
	title={A superconducting quantum simulator based on a photonic-bandgap metamaterial},
	author={Zhang, Xueyue and Kim, Eunjong and Mark, Daniel K and Choi, Soonwon and Painter, Oskar},
	journal={Science},
	volume={379},
	number={6629},
	pages={278--283},
	year={2023},
	publisher={American Association for the Advancement of Science},
        url={https://www.science.org/doi/10.1126/science.ade7651}
}

@article{kohmoto_zero_2007,
  title = {Zero modes and edge states of the honeycomb lattice},
  author = {Kohmoto, Mahito and Hasegawa, Yasumasa},
  journal = {Phys. Rev. B},
  volume = {76},
  issue = {20},
  pages = {205402},
  numpages = {6},
  year = {2007},
  month = {Nov},
  publisher = {American Physical Society},
  doi = {10.1103/PhysRevB.76.205402},
  url={https://journals.aps.org/prb/abstract/10.1103/PhysRevB.76.205402}
}

@article{youssefi2025realization,
  title={Realization of tilted Dirac-like microwave cone in superconducting circuit lattices},
  author={Youssefi, Amir and Motavassal, Ahmad and Kono, Shingo and Jafari, Seyed Akbar and Kippenberg, Tobias J},
  journal={arXiv:2501.10434},
  year={2025},
  url={https://arxiv.org/abs/2501.10434}
}

@article{vool2017introduction,
  title={Introduction to quantum electromagnetic circuits},
  author={Vool, Uri and Devoret, Michel},
  journal={International Journal of Circuit Theory and Applications},
  volume={45},
  number={7},
  pages={897--934},
  year={2017},
  publisher={Wiley Online Library},
  url={https://onlinelibrary.wiley.com/doi/abs/10.1002/cta.2359}
}

@article{youssefi2022topological,
  title={Topological lattices realized in superconducting circuit optomechanics},
  author={Youssefi, Amir and Kono, Shingo and Bancora, Andrea and Chegnizadeh, Mahdi and Pan, Jiahe and Vovk, Tatiana and Kippenberg, Tobias J},
  journal={Nature},
  volume={612},
  number={7941},
  pages={666--672},
  year={2022},
  publisher={Nature Publishing Group UK London},
  url={https://www.nature.com/articles/s41586-022-05367-9}
}

@article{jouanny2025high,
  title={High kinetic inductance cavity arrays for compact band engineering and topology-based disorder meters},
  author={Jouanny, Vincent and Frasca, Simone and Weibel, Vera Jo and Peyruchat, L{\'e}o and Scigliuzzo, Marco and Oppliger, Fabian and De Palma, Franco and Sbroggi{\'o}, Davide and Beaulieu, Guillaume and Zilberberg, Oded and others},
  journal={Nature Communications},
  volume={16},
  number={1},
  pages={3396},
  year={2025},
  publisher={Nature Publishing Group UK London},
  url={https://www.nature.com/articles/s41467-025-58595-8}
}

@article{scigliuzzo_controlling_2022,
	title = {Controlling {Atom}-{Photon} {Bound} {States} in an {Array} of {Josephson}-{Junction} {Resonators}},
	volume = {12},
	url = {https://link.aps.org/doi/10.1103/PhysRevX.12.031036},
	doi = {10.1103/PhysRevX.12.031036},
	number = {3},
	urldate = {2024-02-22},
	journal = {Phys. Rev. X},
	author = {Scigliuzzo, Marco and Calaj{\'o}, Giuseppe and Ciccarello, Francesco and Perez Lozano, Daniel and Bengtsson, Andreas and Scarlino, Pasquale and Wallraff, Andreas and Chang, Darrick and Delsing, Per and Gasparinetti, Simone},
	month = sep,
	year = {2022},
	pages = {031036},
}

@article{amo_tilted_2019,
  title = {Type-III and Tilted Dirac Cones Emerging from Flat Bands in Photonic Orbital Graphene},
  author = {Milicevic, M. and Montambaux, G. and Ozawa, T. and Jamadi, O. and Real, B. and Sagnes, I. and Lemaitre, A. and Le Gratiet, L. and Harouri, A. and Bloch, J. and Amo, A.},
  journal = {Phys. Rev. X},
  volume = {9},
  issue = {3},
  pages = {031010},
  numpages = {11},
  year = {2019},
  month = {Jul},
  publisher = {American Physical Society},
  doi = {10.1103/PhysRevX.9.031010},
  url = {https://link.aps.org/doi/10.1103/PhysRevX.9.031010}
}

@article{bravyi_schriefferwolff_2011,
	title = {Schrieffer–{Wolff} transformation for quantum many-body systems},
	volume = {326},
	issn = {0003-4916},
	url = {https://www.sciencedirect.com/science/article/pii/S0003491611001059},
	doi = {10.1016/j.aop.2011.06.004},
	number = {10},
	urldate = {2024-02-26},
	journal = {Annals of Physics},
	author = {Bravyi, Sergey and DiVincenzo, David P. and Loss, Daniel},
	month = oct,
	year = {2011},
	pages = {2793--2826},
}

@article{bloch_ber_1929,
	title = {{U}ber die {Quantenmechanik} der {Elektronen} in {Kristallgittern}},
	volume = {52},
	issn = {1434-6001, 1434-601X},
	url = {http://link.springer.com/10.1007/BF01339455},
	doi = {10.1007/BF01339455},
	number = {7-8},
	urldate = {2024-02-26},
	journal = {Z. Physik},
	author = {Bloch, Felix},
	month = jul,
	year = {1929},
	pages = {555--600},
}

@book{Girvin_Yang_2019, 
    place={Cambridge}, 
    title={Modern Condensed Matter Physics}, 
    publisher={Cambridge University Press}, 
    author={Girvin, Steven M. and Yang, Kun}, 
    year={2019},
    url={https://www.cambridge.org/highereducation/books/modern-condensed-matter-physics/F0A27AC5DEA8A40EA6EA5D727ED8B14E}
}

@article{underwood2016imaging,
  title={Imaging photon lattice states by scanning defect microscopy},
  author={Underwood, DL and Shanks, WE and Li, Andy CY and Ateshian, Lamia and Koch, Jens and Houck, Andrew A},
  journal={Physical Review X},
  volume={6},
  number={2},
  pages={021044},
  year={2016},
  publisher={APS},
  url={https://journals.aps.org/prx/abstract/10.1103/PhysRevX.6.021044}
}

@article{morvan2021observation,
  title={Observation of topological valley Hall edge states in honeycomb lattices of superconducting microwave resonators},
  author={Morvan, Alexis and F{\'e}chant, Mathieu and Aiello, Gianluca and Gabelli, Julien and Est{\'e}ve, J{\'e}r{\^o}me},
  journal={Optical Materials Express},
  volume={11},
  number={4},
  pages={1224--1233},
  year={2021},
  publisher={Optical Society of America},
  url={https://opg.optica.org/ome/fulltext.cfm?uri=ome-11-4-1224&id=449530}
}

@article{po_fragile_2018,
  title = {Fragile Topology and Wannier Obstructions},
  author = {Po, Hoi Chun and Watanabe, Haruki and Vishwanath, Ashvin},
  journal = {Phys. Rev. Lett.},
  volume = {121},
  issue = {12},
  pages = {126402},
  numpages = {6},
  year = {2018},
  month = {Sep},
  publisher = {American Physical Society},
  doi = {10.1103/PhysRevLett.121.126402},
  url = {https://link.aps.org/doi/10.1103/PhysRevLett.121.126402}
}

@book{Ashcroft:102652,
      author        = "Ashcroft, Neil W and Mermin, N David",
      title         = "{Solid state physics}",
      publisher     = "Holt, Rinehart and Winston",
      address       = "New York, NY",
      year          = "1976",
      url           = "https://cds.cern.ch/record/102652",
}

@article{sanchez_limits_2020,
  title = {Limits of photon-mediated interactions in one-dimensional photonic baths},
  author = {S\'anchez-Burillo, Eduardo and Porras, Diego and Gonz\'alez-Tudela, Alejandro},
  journal = {Phys. Rev. A},
  volume = {102},
  issue = {1},
  pages = {013709},
  numpages = {12},
  year = {2020},
  month = {Jul},
  publisher = {American Physical Society},
  doi = {10.1103/PhysRevA.102.013709},
  url = {https://link.aps.org/doi/10.1103/PhysRevA.102.013709}
}

@article{de_bernardis_light-matter_2021,
	title = {Light-{Matter} {Interactions} in {Synthetic} {Magnetic} {Fields}: {Landau}-{Photon} {Polaritons}},
	volume = {126},
	shorttitle = {Light-{Matter} {Interactions} in {Synthetic} {Magnetic} {Fields}},
	url = {https://link.aps.org/doi/10.1103/PhysRevLett.126.103603},
	doi = {10.1103/PhysRevLett.126.103603},
	number = {10},
	urldate = {2024-02-27},
	journal = {Phys. Rev. Lett.},
	author = {De Bernardis, Daniele and Cian, Ze-Pei and Carusotto, Iacopo and Hafezi, Mohammad and Rabl, Peter},
	month = mar,
	year = {2021},
	pages = {103603},
}

@article{barik2018topological,
  title={A topological quantum optics interface},
  author={Barik, Sabyasachi and Karasahin, Aziz and Flower, Christopher and Cai, Tao and Miyake, Hirokazu and DeGottardi, Wade and Hafezi, Mohammad and Waks, Edo},
  journal={Science},
  volume={359},
  number={6376},
  pages={666--668},
  year={2018},
  publisher={American Association for the Advancement of Science},
  url = {https://www.science.org/doi/10.1126/science.aaq0327}
}

@article{lemonde2019quantum,
  title={Quantum state transfer via acoustic edge states in a 2D optomechanical array},
  author={Lemonde, Marc-Antoine and Peano, Vittorio and Rabl, Peter and Angelakis, Dimitris G},
  journal={New Journal of Physics},
  volume={21},
  number={11},
  pages={113030},
  year={2019},
  publisher={IOP Publishing},
  url={https://iopscience.iop.org/article/10.1088/1367-2630/ab51f5}
}

@article{jalali2020chiral,
  title={Chiral topological photonics with an embedded quantum emitter},
  author={Jalali Mehrabad, Mahmoud and Foster, Andrew P and Dost, Ren{\'e} and Clarke, Edmund and Patil, Pallavi K and Fox, A Mark and Skolnick, Maurice S and Wilson, Luke R},
  journal={Optica},
  volume={7},
  number={12},
  pages={1690--1696},
  year={2020},
  publisher={Optical Society of America},
  url = {https://opg.optica.org/optica/fulltext.cfm?uri=optica-7-12-1690&id=444005}
}

@article{arcari2014near-unity,
  title = {Near-Unity Coupling Efficiency of a Quantum Emitter to a Photonic Crystal Waveguide},
  author = {Arcari, M. and Sollner, I. and Javadi, A. and Lindskov Hansen, S. and Mahmoodian, S. and Liu, J. and Thyrrestrup, H. and Lee, E. H. and Song, J. D. and Stobbe, S. and Lodahl, P.},
  journal = {Phys. Rev. Lett.},
  volume = {113},
  issue = {9},
  pages = {093603},
  numpages = {5},
  year = {2014},
  month = {Aug},
  publisher = {American Physical Society},
  doi = {10.1103/PhysRevLett.113.093603},
  url = {https://link.aps.org/doi/10.1103/PhysRevLett.113.093603}
}

\end{document}